# Chem-SIM: Super-resolution Chemical Imaging via Photothermal Modulation of Structured-Illumination Fluorescence

Dashan Dong[1,2], Danchen Jia[1,2], Xinyan Teng[3], Jianpeng Ao[1,2], George Abu-Aqil[1], Biwen Gao[3], Meng Zhang[1,2], Qing Xia[1,2], Ji-Xin Cheng[1,2,3,4,*]

[1] Department of Electrical and Computer Engineering, Boston University, Boston, Massachusetts 02215, United States;

[2] Photonics Center, Boston University, Boston, Massachusetts 02215, United States;

[3] Department of Chemistry, Boston University, Boston, Massachusetts 02215, United States;

[4] Department of Biomedical Engineering, Boston University, Boston, Massachusetts 02215, United States.

* Correspondence: jxcheng@bu.edu (J.-X.C.)

## Abstract

Structured illumination microscopy (SIM) has attained high spatiotemporal delineation of subcellular architecture, yet offers limited insight into chemical composition. We develop Chem-SIM, a structured-illumination fluorescence detected mid-infrared photothermal microscopy, for super-resolved chemical imaging of microorganisms and mammalian cells. Poisson maximum-likelihood demodulation and spectral normalization across wavenumber recover the weak IR-induced fluorescence intensity change under low photon budgets and convert the fluorescence intensity modulation to chemical fingerprints. Photothermal gating further rejects water backgrounds in aqueous samples, while the IR pump maintains cellular activity at near-physiological temperature. Chem-SIM preserves full vibrational fingerprints, achieves SIM-grade lateral resolution in a high-throughput camera-based format. Here, we show that this platform distinguishes stationary- from log-phase bacteria through chemical content mapping, reports deuterated fatty-acid incorporation in ovarian cancer cells, and resolves lipid-droplet dynamics in live cells, establishing a high-throughput route to super-resolved imaging of organelle chemistry, metabolism, and dynamics.

## Introduction

Organelles partition intracellular processes in space and function, coordinating discrete metabolic pathways and mediating the exchange of materials and information[1, 2]. Recent advances in super-resolution fluorescence microscopy have permitted visualization of individual organelles in living cells at sub-100 nm lateral resolution[3]. Among these modalities, structured-illumination microscopy (SIM) is well suited to live-cell studies because it provides two-fold resolution enhancement with camera-based acquisition with low phototoxicity[4, 5, 6]. By employing multicolor fluorescent tags with SIM, researchers have inferred organelle interactions from their physical contacts[1, 7, 8] and enrich extra dimensions through modalities such as fluorescence lifetime[9], single-molecule photophysics[10], and dipole-orientation analysis[11, 12].

Fluorescence microscopy, however, only visualizes selected chromophores and therefore cannot reveal molecular composition of organelles for function analysis. Advanced vibrational microscopies[13], including coherent Raman scattering[14] and mid-infrared photothermal (MIP) microscopy[15], together with vibration probes[16], provide molecular contrast by probing chemical bond vibrations. Super-resolution coherent Raman imaging via single-pixel detection and structured illumination has been demonstrated on biological specimens[17, 18, 19], whereas the coherent Raman methods require complex ultrafast lasers and remain throughput-limited by point scanning. By contrast, MIP exploits infrared (IR) absorption with cross-sections orders of magnitude larger than those of Raman scattering, uses pump–probe detection to reach sub-micron resolution, and can be implemented in wide-field configurations for high-throughput chemical imaging[20, 21, 22, 23].

In wide-field MIP, a camera detects photothermal modulation of scattering while an IR laser illuminates the full field of view, maximizing IR-energy use[24]; synthetic-aperture computation extends the resolution into three dimensions[25, 26]. Nonetheless, shallow modulation of scattered photons limits its sensitivity[27]. Fluorescence-detected MIP (F-MIP) mitigates this issue by using fluorescent dyes as nanoscale thermometers, increasing modulation depth by two orders of magnitude[28, 29, 30, 31, 32]. Here, the chemical specificity is set by mid-IR absorption of molecular vibrations, while fluorescence provides a sensitive readout of the photothermal response. By integrating F-MIP with structured illumination, the technique can surpass the visible diffraction limit[33]. Yet, strong water absorption[34] and limited fluorescence photon budgets[28, 31] in living specimens have thus far hindered super-resolution wide-field photothermal imaging in aqueous environments, despite proof-of-concept demonstrations on dried polymer beads[33].

Here, we present Chem-SIM, a wide-field MIP-modulated structured-illumination fluorescence platform for super-resolution chemical imaging in both fixed and live cells. By integrating sinusoidal illumination with Poisson maximum-likelihood demodulation, Chem-SIM achieves ~2× lateral resolution while boosting sensitivity over previously reported wide-field F-MIP. Our approach preserves full vibrational fingerprints under photon-limited conditions and suppresses water background through photothermal relaxation gating. Notably, the IR pump maintains a near-physiological temperature during acquisition to help sustain cellular activity. With such capacities, Chem-SIM enables fingerprint-level readout of metabolism and chemistry in single bacterial and single organelles. Chem-SIM requires only minimal optical modifications to a standard wide-field microscope, and provides a platform for high-throughput mapping of organelle chemistry, metabolism, and dynamics.

# Results

## Principle and implementation of Chem-SIM

Chem-SIM integrates mid-infrared (mid-IR) photothermal excitation with structured-illumination microscopy to deliver super-resolution, wide-field chemical imaging. As shown in **Figure 1a**, the system adopts an upright wide-field layout optimized for calcium fluoride ($CaF_2$) substrates. A tunable 50 kHz mid-IR quantum cascade laser (QCL) excites molecular vibrations. A nanosecond-pulsed visible laser is patterned by a digital micromirror device (DMD) to encode high-spatial-frequency content beyond the diffraction limit. Photothermal-induced fluorescence modulation is acquired in alternating "hot" (IR on) and "cold" (IR off) frames that are pulse-synchronized via electronic gating (**Figure 1b**). Hundreds of pump–probe cycles are averaged per frame to approach shot-noise-limited fluorescence detection by an sCMOS camera. Full details of the optical and electronic components are provided in the *Methods*.

For each Chem-SIM acquisition, three pattern orientations and three phase shifts yield nine hot and nine cold images. These raw frames are reconstructed using a joint RL–SIM algorithm, which preserves quantitative modulation contrast (detailed in **Supplementary Note 1** and **Supplementary Figure 1**), and the resulting hot-SIM and cold-SIM images are then subtracted pixel-wise to extract the thermally modulated signal (**Figure 1c**). For hyperspectral imaging, the mid-IR pump is stepped from 900–1798 $cm^{-1}$ in 1 $cm^{-1}$ increments; at each wavenumber we record the full 18-frame sequence at speed of 40 fps with an effective exposure of 23.1 ms per frame, giving a total acquisition time of ~585 s for the entire stack (detailed in **Supplementary Note 2**). As shown in **Figure 1d**, cold-SIM exhibits ~20% photobleaching over the scan, whereas hot-SIM additionally encodes wavelength-dependent IR absorption. Here, the raw modulation-depth spectrum is defined as the relative IR-induced fluorescence change referenced to the cold-state fluorescence. After normalization by the cold-SIM bleaching curve, the cold–hot subtraction reveals water-vapor absorption lines along the optical path, consistent with the on-sample IR power spectrum (**Figure 1e**). Finally, a weighted least-squares normalization of this modulation-depth spectrum by the measured on-sample IR power both equalizes wavenumber-dependent pump power and effectively suppresses water-vapor absorption lines, producing the Chem-SIM absorption spectrum (**Figure 1f**) that closely matches the normalized reference Fourier transform infrared (FTIR) spectrum. Full details of spectrum normalization are provided in the *Methods*, **Supplementary Note 3,** and **Supplementary Figure 2**.

## High-Quality Chem-SIM by Poisson Maximum-Likelihood Demodulation

Despite the high-quality area-averaged Chem-SIM spectra (**Figure 1f**), single-pixel Cold–Hot images remain dominated by noise Because mid-IR absorption induces only a weak local fluorescence modulation ($< 10\%$ on average). As shown in **Figure 2a**, the raw hot and cold frames have sufficient signal-to-noise ratio (SNR) to reveal the structured-illumination fringes; however, their cold–hot subtraction is noise-amplifying and fails to recover these patterns. This follows from photon statistics, differencing two Poisson-distributed measurements produces a Skellam-distributed signal whose variance scales with the sum of the counts, degrading SNR when the modulation depth is shallow. Even near saturation of the cold channel, the differential signal remains compressed into a narrow range, where noise amplification and brightness-modulation coupling can obscure true chemical contrast (**Figure 2b**).

To address these limitations, we developed a likelihood-based demodulation strategy that exploits the full hot/cold photon statistics together with the SIM forward model. Specifically, we introduce a Poisson maximum-likelihood demodulation (P-MLD) framework[35] that jointly models the unmodulated fluorescence and the IR-induced modulation under structured illumination (**Figure. 2c**). Let $x \geqslant 0$ denote the baseline (cold) fluorescence map, and let $0 \leq m \leq 1$ denote a dimensionless modulation map such that the Hot fluorescence map is given by the element-wise product $m \odot x$. The photothermal modulation image used as the chemical-contrast channel is defined as the cold–hot difference $z \equiv x - m \odot x$. For each structured-illumination pattern, we adopt a standard SIM forward model that multiplies $x$ by the pattern, convolves with the system point-spread function, and adds a slowly varying background term. Each pixel in all hot and cold frames is treated as a Poisson random variable whose mean is given by this forward model, and P-MLD seeks nonnegative $x$, $m$, and background that maximize the joint Poisson likelihood of all frames; we solve the resulting inverse problem using a nonnegative Richardson–Lucy–style multiplicative algorithm (detailed in *Methods* and **Supplementary Note 4**).

**Figure 2d-g** benchmarks P-MLD against conventional differencing on the same dataset. Applied to a single raw structured-illumination hot/cold pair, P-MLD suppresses spurious negatives, enhances local SNR, and cleanly recovers the fringe pattern (**Figure 2d**). Using the full 3-orientation × 3-phase dataset, SIM-based P-MLD produces a modulation image with improved effective resolution and recovery of finer features, enabled an SNR improvement of ~20 dB

compared with pixel-wise cold-hot subtraction (**Figure 2e** and detailed in **Supplementary Note 5**). At the single-pixel level, P-MLD markedly reduces spectral noise (**Figure 2f**). After applying spectral normalization described in the previous section, the recovered high-SNR vibrational spectrum closely matches the area-averaged Chem-SIM spectrum (**Figure 2g**). Together, these results demonstrate that P-MLD enables high-fidelity Chem-SIM mapping at each wavenumber without increasing the photon budget or compromising spectral resolution.

## Spatial and spectroscopic mapping of single *Staphylococcus aureus*

We first applied Chem-SIM to *Staphylococcus aureus* (*S. aureus*) labeled with rhodamine 6G and mounted on $CaF_2$ coverslips (**Figure 3**). For a fair comparison under identical photon budgets, nine patterned raw frames were averaged to synthesize the equivalent cold/hot wide-field fluorescence (WF-FL) images, and an F-MIP image was obtained by cold–hot subtraction. The Chem-SIM images are computed from the same raw data using the P-MLD algorithm with structured illumination. In **Figure 3a-d**, as expected for SIM, WF-FL provides only a blurred outline of individual cocci, whereas SIM resolves the granular cell-wall morphology across the same field. Line profiles and Fourier ring correlation (FRC) analysis[36] (detailed in **Supplementary Note 6**) confirm an almost twofold gain in lateral resolution for SIM relative to WF-FL.

Probing the amide-I band (1650 $cm^{-1}$) produces chemical contrast in both F-MIP and Chem-SIM that is dominated by the mid-IR focus profile (**Figure 3e**). In the F-MIP images, however, the cold–hot subtraction leaves substantial high-frequency noise, individual cocci remain only weakly separated, and the apparent fine texture is largely noise-driven. By contrast, with P-MLD reconstruction, Chem-SIM recovers individual cells and subcellular texture with markedly enhanced SNR (**Figure 3f**). Due to additional noise from subtraction, the F-MIP images exhibit high-frequency fluctuations that do not follow the underlying fluorescence distribution (**Figure 3g**), and FRC analysis shows that this noise actually degrades the effective resolution from ~421 nm for WF-FL to ~708 nm for F-MIP (**Figure 3h**).

At the nucleic-acid band with lower modulation (1080 $cm^{-1}$), both F-MIP and Chem-SIM show markedly reduced signal levels (**Figure 3i**). Zoomed views (**Figure 3j**) and line profiles (**Figure 3k**) show that Chem-SIM retains lower noise and a spatial distribution that more closely tracks the fluorescence structure than F-MIP. As anticipated from the FRC benchmark of 214 nm (**Figure 3,l**), Chem-SIM at both 1650 and 1080 $cm^{-1}$ achieves SIM-like structural detail while preserving mid-IR spectral specificity. On the contrary, although F-MIP shows good contrast at strongly absorbing bands, it rapidly loses spatial discriminability at lower-modulation bands when based solely on cold–hot subtraction.

*S. aureus* is responsible for over 50% of human skin infections. Distinguishing between log and stationary phases of *S. aureus* is significant due to shifts in the bacterial virulence factor expression, metabolism, and antibiotic susceptibility. With much enhanced contrast and resolution in Chem-SIM, we next compared stationary-phase and log-phase populations at the single-coccus level. Chem-SIM maps at 1650 $cm^{-1}$ (**Figure 3m**) reveal heterogeneous cellular chemistry, and single-bacterium–averaged Chem-SIM spectra (mean ± s.e.m.) capture clear population-level trends (**Figure 3n**). During rapid growth, nucleic-acid synthesis elevates the $PO_2^-$ band over amide-I band, as reflected in an increased 1080/1650 $cm^{-1}$ intensity ratio for log-phase bacteria (**Figure 3o**). Together, these results show the potential of Chem-SIM not only for the detection of *S. aureus*, but also characterization of its metabolic activity.

## Background Suppressed Chem-SIM Imaging in Aqueous Environment

Applying our Hot/Cold SIM–based Chem-SIM to HeLa cells in buffer, a strong signal for the ester C=O vibration at 1744 $cm^{-1}$ was expected (**Figure 4a**). Instead, hyperspectral Chem-SIM shows that Cold-SIM traces follow normal photobleaching, whereas Hot-SIM modulation closely tracks the IR laser power spectrum, and even power-normalized spectra increase near the water absorption band at 1650 $cm^{-1}$ (**Figure 4b**). Consistent with this, we observe distinct photothermal relaxation kinetics at the lipid carbonyl (C=O) band and at water-dominated bands (**Supplementary Figure 3**). Finite-element simulations support these observations. Under wide-field illumination, repeated mid-IR pulse trains establish a quasi-steady "heat zone" in the aqueous medium just above the $CaF_2$ substrate, slightly beyond the nominal penetration depth, which equilibrates on the millisecond timescale[34]. Individual pump pulses still induce fast transient excursions on top of this background. Because bulk water is thermally coupled to this extended heat zone, it cools more slowly than small, thermally confined inclusions such as lipid droplets or 100-nm PMMA beads, leading to a delayed thermal transient at water-dominated bands (**Supplementary Note 7, Supplementary Figure 4 & Supplementary Movie 1**). Such thermal dynamics difference raises the opportunity of extracting small signals from the water absorption background in widefield photothermal imaging of aquatic samples.

Leveraging the slowly varying feature of water background, we introduce photothermal relaxation (PR) gating for wide-field photothermal imaging. Conceptually (**Figure 4c**), the pump pulse creates a fast temperature rise in the absorber (lipid) and a slower, spatially diffuse response in water. PR gating acquires two frames at two different probe delays. Practically, Delay 1 is chosen near the prompt maximum of the absorber-dominated transient, whereas Delay 2 is chosen in a later water-dominated regime where the target and background responses are most distinct. Their difference suppresses the water background while retaining target-specific contrast (**Supplementary Note 8 & Supplementary Figure 5-6**). PR gating is implemented electronically by shifting the mid-IR trigger relative to the probe, without additional boxcar hardware and without sacrificing wide-field throughput, and the two-delay frames are integrated into the P-MLD framework to yield PR gated Chem-SIM reconstructions that remain non-negative and resolution-enhanced.

Applied to Lipi-Red–stained HeLa cells, SIM provides the expected gain over WF-FL, yielding higher contrast and an approximately twofold improvement in lateral resolution (**Figure 4d–f**). With PR gating, the aqueous background is strongly suppressed in both F-MIP and Chem-SIM, isolating lipid droplets with high local contrast (**Figure 4g, h**). FRC analysis further shows that PR-gated Chem-SIM achieves ~4× improved effective spatial detail relative to PR-gated F-MIP, reflecting the combined benefits of higher SNR and improved recovery of high-frequency information under P-MLD (**Figure 4f, i**). PR gated Chem-SIM spectra (**Figure 4j**) from the eight individual droplet marked in **Figure 4h** consistently show a dominant ester C=O peak at 1744 $cm^{-1}$, together with other lipid bands[37] showing robust fingerprint-level specificity. The peak around 1650 $cm^{-1}$ is assigned to residual water background. Together, these results demonstrate background-suppressed, super-resolved chemical imaging of lipid droplets in aqueous media without sacrificing photon budget or spectral fidelity.

## Chem-SIM Unveils Lipid Chemistry in Cancer Cells

Ovarian cancer cells are known to accumulate lipids and gain aggressiveness in a lipid-rich environment[38, 39]. To explore the potential of Chem-SIM for imaging lipid chemistry inside ovarian cancer cells, OVCAR5 cells were first incubated in lipid-deficient medium and then treated with fully-deuterated palmitic acid-$d_{31}$ (PA-$d_{31}$) for different time periods. Chem-SIM spectra cover both the fingerprint region and the C–D "silent window" (**Figure 4k & Supplementary Figure 7**). After cell fixation with formaldehyde, hyperspectral Chem-SIM imaging was carried out. In the fingerprint region, lipid droplets display various signatures, including a prominent ester C=O band at 1744 $cm^{-1}$, $CH_2$ vibrations near 1464 $cm^{-1}$, and $CH_3$ vibrations near 1377 $cm^{-1}$. The band at ~1164 $cm^{-1}$ is consistent with C–O–C symmetric stretching in the glycerol backbone. In cells treated with lipid-deficient medium, an additional band appears near 1100 $cm^{-1}$, attributable to C–O stretching in C–OH groups of diacylglycerol (DAG) or monoacylglycerol (MAG) species produced by lipase-mediated triacylglycerol (TAG) hydrolysis under fatty-acid deprivation[40]. The corresponding TAG, DAG structures and synthesis are shown in **Supplementary Figure 8**.

With increasing PA-$d_{31}$ incubation time, strong symmetric and antisymmetric C–D vibration bands emerge and grow in the silent window, indicating progressive incorporation of exogenous deuterated fatty acids into lipid droplets as a protective mechanism against lipotoxic stress[41]. In particular, compared with the PA-free cells, the antisymmetric C–O–C band near 1255 $cm^{-1}$ is markedly enhanced, consistent with an increased population of asymmetric TAG species arising from deuterated acyl chains[40]. By contrast, the C-O peak near 1100 $cm^{-1}$ is much reduced. These systematic spectral changes in both silent and fingerprint windows report on strong TAG biosynthesis activity upon external palmitic acid treatment.

## Chem-SIM allows dynamic imaging of live cells

A practical challenge for Chem-SIM imaging of live cells is cumulative heating from repeated mid-IR excitation, which can perturb cell physiology. In our implementation, live-cell Chem-SIM is enabled by actively constraining the thermal load in three ways. First, the use of a water-dipping objective and a relatively large imaging dish, together with water's high heat capacity, facilitates rapid thermal exchange between the field of view and the surrounding medium. Second, under fixed IR-pump conditions, our nanosecond probe timing is positioned close to the peak of the photothermal transient, maximizing modulation depth for a given average power. Third, each Chem-SIM frame integrates over ~4,500 pump pulses, providing high sensitivity to small temperature changes while averaging out laser and thermal noise. Fluorescence thermometry with Lipi-Red (**Supplementary Note 9** and **Supplementary Figures 9–10**) shows that the mean medium temperature during imaging scales with the mid-IR average power. At the lipid C=O absorption peak, the surrounding medium equilibrates at ~34 °C, while individual mid-IR pulse trains drive rapid excursions to ~38 °C that relax back to baseline within the interpulse interval.

With optimized heat control, we applied PR-gated Chem-SIM to track lipid droplets in live OVCAR5 cells over tens of minutes (**Figure 5a & Supplementary Movie 2**). Unlike the full hyperspectral acquisitions used for fingerprint-region characterization, the live-cell experiments employ selected diagnostic bands to prioritize temporal resolution for time-lapse imaging. Specifically, each time point consisted of sequential acquisitions at 1744 $cm^{-1}$ and 2196 $cm^{-1}$. SIM Lipi-Red images define the droplet morphology, whereas Chem-SIM maps at 1744 $cm^{-1}$ (ester C=O) and 2196 $cm^{-1}$ (C–D) report the endogenous lipid and deuterated TAG pools, respectively. Under IR-ON conditions, both C=O and C–D contrast are observed. The IR-OFF controls show a negligible Chem-SIM signal.

To examine the impact of IR exaction on cell physiology, we performed single-droplet tracking (**Figure 5b**), which yields time series that are well described by an anomalous-diffusion model[42, 43]. From this data, we extracted the mean-squared displacement (MSD), anomalous exponent $\alpha$, and generalized diffusion coefficient $D$. Box-plot summaries (**Figure 5c–d**) show that SIM-based wide-field imaging (IR-OFF), performed at room temperature of 20°C, preserves the expected super-diffusive motion of lipid droplets. Interestingly, IR illumination produces little change over the thermal-driven diffusion factor $D$ (**Figure 5e**), indicating that Chem-SIM can monitor lipid-droplet dynamics wihle preserving motility under the tested imaging conditions. Together, these results demonstrate that Chem-SIM allows background-suppressed, super-resolved, and dynamically compatible chemical imaging of living cells under the experimental conditions examined here.

# Discussion

By mid-IR photothermal modulation of structured-illumination fluorescence and Poisson likelihood demodulation, Chem-SIM is developed to deliver chemical maps of single bacteria and single organelles beyond the visible beam diffraction limit. Bridging super-resolution fluorescence with vibrational spectroscopy, Chem-SIM opens a practical route to map organelle chemistry and microorganism metabolism—an enabling capability for cell biology, microbiology, and drug discovery.

Chem-SIM occupies a distinct niche among existing imaging modalities. Advanced fluorescence microscopy provides high spatial resolution but limited chemical specificity, whereas coherent Raman and other vibrational imaging methods offer chemical contrast but often remain constrained by instrumentation complexity or imaging throughput. Building on wide-field MIP and F-MIP, Chem-SIM combines structured illumination, Poisson maximum-likelihood demodulation, and photothermal relaxation gating to enable fingerprint-level chemical imaging with improved spatial resolution, background suppression, and compatibility with aqueous samples.A central challenge in wide-field photothermal imaging is the Skellam barrier where the difference of two Poisson measurements has variance that scales with the total detected fluorescence intensity. Such shallow modulations are easily drowned by shot noise and dynamic-range compression. P-MLD addresses this challenge by fitting the full Hot/Cold photon statistics under the SIM forward model, recovering modulation amplitudes without introducing negative artifacts and with markedly lower noise floors. In a single *S. aureus*, the approach improves the FRC resolution from ~708 nm (F-MIP) to ~214 nm while preserving spectral features that match FTIR references. Conceptually, P-MLD is modality-agnostic and could be adapted to other pump–probe imaging schemes that suffer from similar brightness–modulation coupling. Accordingly, the observed FRC improvement likely reflects both noise suppression and improved recovery of high-frequency chemical contrast under the model-based reconstruction, rather than solely an intrinsic optical-resolution increase. The recovered Chem-SIM modulation maps should therefore be interpreted as model-supported chemical contrast inferred from the joint Hot/Cold measurements, rather than as fluorescence morphology transferred into the chemical channel by the estimator.

In aqueous media, water absorption of mid-IR light overwhelms weak biomolecular signals and limits the modulation depth. Our PR gating exploits the temporal asymmetry between fast absorber heating/cooling and the slower, axially conducted water response. The two probe delays are chosen to sample the lipid-dominated peak and the water-dominated tail, and the difference cancels the water background while retaining target-specific contrast. While the principle of PR gating is general, the optimal delay pair is an empirically selected operating condition that depends on the sample geometry, thermal environment, and imaging target, rather than a universally fixed parameter. Because water's thermal response is strongly axial, effective optical sectioning becomes critical; SIM provides near-confocal sectioning in a camera-based format, making PR-gated chem-SIM a powerful tool for chemical imaging in liquid environments. Importantly, PR gating already suppresses background at the wide-field F-MIP level, indicating its broader applicability to synchronized wide-field photothermal imaging.

Chem-SIM provides the full mid-IR fingerprint spectrum that tells the chemical content of fluorescence-labeled bacteria or organelles. Such information lies beyond the reach of super-resolution fluorescence microscopy but is essential for organelle functional analysis. In Chem-SIM, the fluorescent reporter plays a dual role as a structural marker and a nanoscale thermometer. Chemical specificity arises from mid-IR absorption by molecular vibrations, whereas fluorescence serves as the detection channel that reports the IR-induced photothermal modulation. Pulsed mid-IR absorption encodes chemical bond information to the fluorescence contrast. Such capability turns fluorescence-resolved morphology into chemically interpretable maps, which is essential for organelle functional analysis and metabolic phenotyping.

We have shown that Chem-SIM images inherit the spatial resolution of the fluorescence channel. In F-MIP with a continuous wave probe, the point spread function (PSF) is set by the fluorescence PSF convolved with a thermal diffusion kernel accumulated over the detection window. In this work, our nanosecond-pulsed probe gates the signal predominantly during the prompt heating transient, and largely rejects the later, diffusion-driven relaxation. In PR-gated chem-SIM, the same transient selection strategy suppresses the more delocalized aqueous contribution in liquid, producing chemical maps at the SIM resolution in liquid environments.

We would point out several limitations of our technology. First, Chem-SIM relies on fluorescent reporters; the photon budget, photobleaching, and dye distribution can influence SNR and the interpretability of chemical maps. Our analysis and experiments indicate that P-MLD mitigates coupling between brightness and modulation, but absolute quantitation still requires careful control of labeling density and excitation conditions. Second, PR gating relies on differences in thermal transients that can be size- and shape-dependent; heterogeneity in droplet dimensions, morphology, or local thermal boundary conditions may bias the relative weighting of delays and should be characterized when comparing populations. Finally, spectral fidelity depends on accurate power normalization and suppression of atmospheric water lines; our weighted least-squares normalization is robust across reasonable parameter ranges, but seasonal humidity fluctuations necessitate periodic calibration of the power spectrum (or purging of the beam path) to stabilize baseline and line depths.

Several extensions are natural. Like 3D SIM[44, 45], 3D Chem-SIM could be realized by combining optical sectioning with depth-dependent delay selection in PR gating. In addition, by adaptive acquisition, i.e., tuning the IR wavenumber while performing time-lapse imaging, Chem-SIM could track organelle chemistry dynamically while minimizing dose and acquisition time.

# Methods

## Instrumentation

Chem-SIM was implemented on a lab-built upright wide-field microscope. A tunable mid-infrared QCL (Daylight Solutions, MIRcat) served as the pump, and a 520 nm nanosecond pulsed laser diode (Thorlabs, NPL52C) provided fluorescence excitation. Constrained by the probe repetition rate, the pump–probe sequence operated at 50 kHz. To maximize photothermal heating, the pump pulse width was set to 500 ns, and the pump–probe delay was tuned to the peak of the fluorescence-modulation response (**Supplementary Figure 11**).

**Pump path.** The mid-IR beam was focused onto $CaF_2$ substrates with a parabolic mirror (Thorlabs, MPD019-M03), producing an on-sample spot of approximately 50 μm diameter.

**Probe/illumination path.** The 520 nm probe was coupled through a fiber despeckler (Newport, F-DS-AFS105-FC/PC) via a focusing lens (L7, Thorlabs C260TMD-A) to reduce spatial coherence. After 200× beam expansion (L5–L6), the beam impinged on a digital micromirror device (DMD; Texas Instruments, DLP9500) at the blaze angle, carrying binary structured-illumination patterns. The patterned beam was Fourier-filtered with a lens (L4, f = 500 mm) and a pinhole mask (hexagonal aperture array) to pass the ±1 diffraction orders. A relay (L2 f = 180 mm, L3 f = 250 mm) imaged the pattern to the sample through either a Nikon TUPlanFluor, NA 0.8 Air or Olympus LUMFLN, NA 1.1 Water objective. Epi-fluorescence was separated by a filter set (Ex 520/40, DM 550 LP, Em 550 LP), relayed by a tube lens (Thorlabs, TTL200), and recorded on an sCMOS camera (Andor Zyla 5.5) to form wide-field images.

**Synchronization and timing.** A digital delay pulse generator (Quantum Composers, Emerald 9254) provided the 50 kHz master clock and programmable trigger signals. A custom control program computed and issued the appropriate gated signals to the QCL, probe laser, and sCMOS exposure based on the selected camera integration time. The DMD was triggered by the camera's global-exposure output to achieve an effective global shutter when running camera at its high-sensitivity rolling mode; its latency relative to the master clock was measured and actively compensated, ensuring the same number of pump–probe cycles per Hot/Cold frame. Both the pump–probe delay and the PR window were implemented by adjusting the QCL trigger delay relative to the probe in practice, thereby maintaining identical fluorescence-detection conditions across all acquisitions. For PR gating, the delay pair was selected empirically from measured transient responses at a target-dominated band and a water-dominated band: Delay 1 was set near the prompt target maximum, and Delay 2 was set later where the two transients were most distinct. Accordingly, the selected PR window should be interpreted as system- and sample-dependent, rather than as a universal parameter set. This electronic implementation makes PR gating readily transferable to other synchronized wide-field F-MIP systems without additional optical boxcar detection hardware.

## Spectroscopic Calibration and Normalization

**Least squares spectral normalization.** To compensate mid-IR power variation (including water-vapor lines) and remove baseline offsets while preserving sharp spectral features in 1D MIP/Chem-SIM spectra, we estimate a nonnegative normalized spectrum $u \geq 0$ and a scalar baseline $b$ by solving a weighted least-squares problem with TV regularization

$$\min_{u\geq 0,b}\left\{\sum_{i}(w_i u_i - y_i - b)^2 + \lambda \mathrm{TV}(u) + \mu|b|\right\},$$

on the wavenumber grid $\nu_i$. Where $y_i$ is the measured raw spectrum and $w_i = p_i/\max_j p_j$ are weights from the on-sample IR power spectrum $p_i$. Here $\mathrm{TV}(u)$ denotes the 1D total-variation semi-norm (absolute finite differences). This formulation is equivalent to a stabilized division by the power profile ($u \approx y/w$), while denoising edges and removing a constant offset. Solver detailed in **Supplementary Note 3.**

**Across-dataset spectral comparison.** For comparing multiple Chem-SIM/F-MIP spectra, area normalization (divide by the mean value over the full spectrum) is additionally applied to equalize overall intensity and emphasize relative band difference.

**FTIR spectral normalization.** Attenuated total reflection FTIR (ATR-FTIR, Thermo Scientific) spectra were exported as absorbance and converted to absorptance, then energy-weighted by multiplying by wavenumber to match the Chem-SIM/F-MIP energy-absorption convention, resampled onto the Chem-SIM/F-MIP wavenumber grid, and peak-normalized before overlay and comparison. Note: This procedure harmonizes dimensionality for shape comparison; absolute scaling in ATR depends on penetration depth and contact conditions and is not interpreted.

## Poisson maximum-likelihood demodulation (P-MLD)

As mentioned in the main text, the Cold-state image corresponds to the baseline fluorescence map $x_{\mathrm{Cold}} = x$, and the Hot fluorescence map is given by $x_{\mathrm{Hot}} = m \odot x$. For the $k$-th illumination pattern $s_k$ and point-spread-function (PSF) operator $h$, we model the imaging system by the linear operator

$$A_k(x) = h * (s_k \odot x) + b,$$

where $b \geq 0$ represents a slowly varying out-of-focus/background component added after convolution and $*$ denotes convolution. The expected photon counts (Poisson means) for the Cold/Hot frames are

$$\mu_k^{\mathrm{Cold}} = A_k(x), \;\; \mu_k^{\mathrm{Hot}} = A_k(m \odot x).$$

The photothermal modulation image used as the chemical-contrast channel is defined as $z \equiv x - m \odot x$.

For each pattern $k = 1, \cdots, 9$ (3 orientations × 3 phases) and pixel $j$, the observed counts are $I_{k,j}^{\mathrm{Cold}}, I_{k,j}^{\mathrm{Hot}}$. Under shot-noise–limited detection, each pixel/frame is modeled as an independent Poisson random variable with mean $\mu_{k,j}^{\mathrm{Cold}}$ or $\mu_{k,j}^{\mathrm{Hot}}$, with probability mass function

$$p(I|\mu) = \frac{e^{-\mu}\mu^I}{\Gamma(I+1)}.$$

P-MLD jointly estimates $z$ (equivalently $x$ and $m$) by maximizing the joint Poisson log-likelihood[46]:

$$\mathcal{L}(x, m) = \log\left[\prod_{k}^{9}\prod_{j}^{w\times h} p\left(I_k^{\mathrm{Cold}}|\mu_k^{\mathrm{Cold}}\right) p\left(I_k^{\mathrm{Hot}}|\mu_k^{\mathrm{Hot}}\right)\right]$$

Where $j$ indexes pixels, $w \times h$ is the image size, and $p(I|\mu)$ denotes the Poisson probability mass function. In this formulation, the baseline fluorescence map and the IR-induced modulation are estimated jointly but remain distinct unknowns. The reconstruction therefore does not constrain the modulation map to follow the fluorescence morphology; rather, modulation is supported only when it consistently explains the differential Hot/Cold measurements across all structured-illumination patterns. Algorithmic details are provided in **Supplementary Note 4**.

## Sample preparation

**Substrates.** Double-side–polished calcium fluoride ($CaF_2$) disks (Crystran; 1 mm thick) were used as IR-transparent substrates for all samples. Disks were bonded with PMMA (polymethyl methacrylate) adhesive to the bottom aperture of a confocal imaging dish to form the observation window, then rinsed in 70% ethanol, washed with deionized water, and air-dried in a dust-free container.

**HeLa and OVCAR5 cell culture and Lipi-Red labeling.** HeLa cells were purchased from ATCC, and OVCAR5 cells were purchased from Sigma-Aldrich. Cells were cultured under standard conditions (37 °C, 5% $CO_2$) in their recommended media (e.g., DMEM for HeLa, RPMI-1640 for OVCAR-5) supplemented with 10% FBS (v/v) and 1% penicillin–streptomycin (v/v). For imaging, cells were seeded onto sterilized $CaF_2$ disks and cultured overnight for attachment. For fatty-acid treatment experiments, OVCAR5 cells were incubated with BSA-complexed PA-$d_{31}$ in de-lipid culture medium to promote lipid uptake and incorporation. PA-$d_{31}$ was first dissolved to 50 mM stock solution in DMSO and then diluted in de-lipid RPMI-1640 medium to 50 μM as the final concentration. The blank control group was cultured in the same medium condition without PA supply. After treatment, cells were washed 3× with pre-warmed PBS before staining. Lipid droplets were stained with Lipi-Red (Dojindo, 9 μmol $L^{-1}$, de-lipid medium) for 30 min at 37 °C, protected from light. After staining, cells were washed 3× with pre-warmed PBS. For live-cell Chem-SIM imaging of OVCAR5 cells, after staining and washing, the $CaF_2$ disk was transferred to a $CO_2$-independent live-cell imaging buffer (phenol red–free, HEPES-buffered medium) and allowed to equilibrate on the microscope stage for 30 min at ~20 °C room temperature before imaging (typically ~15 min per sample). For fixed-cell imaging, cells were then fixed in 4% paraformaldehyde for 30 min at room temperature, washed 3× with PBS, before imaging. For hyperspectral Chem-SIM of lipid chemistry under fatty-acid treatment, OVCAR5 cells were fixed after PA-d31 (or PA-free) incubation and subsequent washing and then imaged following the same staining and handling steps unless otherwise specified.

**R6G-labeled *E. coli* and *S. aureus*.** Frozen stocks of *Escherichia coli (E. coli)* BW25113 and *Staphylococcus aureus (S. aureus) ATCC6538* stored at –80 °C were streaked onto LB agar plates and incubated overnight at 37 °C. A single colony was then picked and inoculated into 3 mL of LB broth, followed by incubation at 37 °C for 3 hours to reach the logarithmic growth phase and 12 hours to reach the stationary phase. Bacterial cells were pelleted (5,000 x g, 5 min) and washed 3 times with PBS. The bacterial pellets were labeled with Rhodamine-6G (R6G) by incubating in 10 μM dye (in PBS) for 30 min at room temperature, protected from light, followed by 3 sterile water washes to reduce salt residues. A 3–10 μL aliquot was deposited onto a clean $CaF_2$ disk, gently spread to form a thin film, and air-dried for 10-15 min at room temperature, in the dark. Dried specimens were stored desiccated and imaged within several days.

## Imaging data analysis and statistics

Imaging data were acquired and reconstructed with custom MATLAB (Mathworks) scripts. Hyperspectral Chem-SIM data were further processed and analyzed in ImageJ (Fiji). Lipid-droplet dynamic tracking was performed automatically using the TrackMate plugin (ImageJ/Fiji). Graphing, statistical summaries, and figure preparation were completed in MATLAB, OriginPro (OriginLab) and Adobe Illustrator.

Unless otherwise stated, quantitative results are reported as mean ± s.e.m. from $n$ individual cocci or lipid droplets within a signle field of view (FOV), as indicated in the corresponding figure legends. For log-phase/stationary-phase analyses and lipid-droplet experiments, experiments were independently repeated twice, and the quantitative results shown were obtained from one representative FOV from one independent experimental repeat unless otherwise stated. Repeated acquisitions from the same sample, including measurements from different FOVs of the same sample, were not counted as independent experimental repeats.

Statistical analyses were performed in OriginPro and MATLAB. SNR comparisons were performed using the Wilcoxon signed-rank test, log-stationary phase analyses were performed using an unpaired two-sample t-test, and MSD-based lipid-droplet motion analyses were performed using the Mann–Whitney U test (ranksum in MATLAB). All statistical tests were two-sided unless otherwise stated. No covariates were included in the statistical analyses; comparisons were performed directly between the indicated experimental groups or conditions. No formal tests of normality were performed. No adjustment for multiple comparisons was applied unless otherwise stated. Statistical significance was annotated as: *, $p < 0.05$; **, $p < 0.01$; ***, $p < 0.001$; ****, $p < 0.0001$, exact values are provided in the corresponding figure legends.

## Data availability

The open-source code associated with this study includes the necessary datasets required to reproduce the main workflows. Source data for the figures are provided with this paper. Additional datasets generated and analyzed during the current study are available from the corresponding author upon reasonable request.

## Code availability

The weighted least-squares spectral normalization tool and P-MLD demodulation code are available as open-source software via a Code Ocean capsule at https://codeocean.com/capsule/1545326.

## Acknowledgements

This work is supported by NIH grants R35GM136223, R01AI141439, R33CA287046, and by grant number 2023-321163 from the Chan Zuckerberg Initiative DAF, an advised fund of Silicon Valley Community Foundation.

## Author information

### Authors and Affiliations

**Department of Electrical and Computer Engineering, Boston University, Boston, USA**

Dashan Dong, Danchen Jia, Jianpeng Ao, George Abu-Aqil, Meng Zhang, Qing Xia and Ji-Xin Cheng

**Photonics Center, Boston University, Boston, USA**

Dashan Dong, Danchen Jia, Jianpeng Ao, Meng Zhang, Qing Xia and Ji-Xin Cheng

**Department of Chemistry, Boston University, Boston, USA**

Xinyan Teng, Biwen Gao, and Ji-Xin Cheng

**Department of Biomedical Engineering, Boston University, Boston, USA**

Ji-Xin Cheng

## Contributions

D.D. and J.X.C. conceived the project. J.X.C. supervised the research and revised the manuscript. D.D. designed and built the instrumentation, wrote the control software, developed the reconstruction algorithms, and performed all experiments. D.C.J. assisted in the data acquisition and hyperspectral chemical imaging. D.C.J., X.Y.T., J.P.A., and Q.X. conducted substantive discussions on data analysis. X.Y.T., J.P.A., and B.W.G prepared and provided the cell samples, while G.A.-A. and M.Z. prepared and provided the bacterial samples. All authors discussed the results and approved the final manuscript.

## Corresponding author

Ji-Xin Cheng, jxcheng@bu.edu

# Ethics declarations

## Competing interests

JXC declare financial interest with Photothermal Spectroscopy Corp, which did not fund this work. Other authors declare no competing interests.

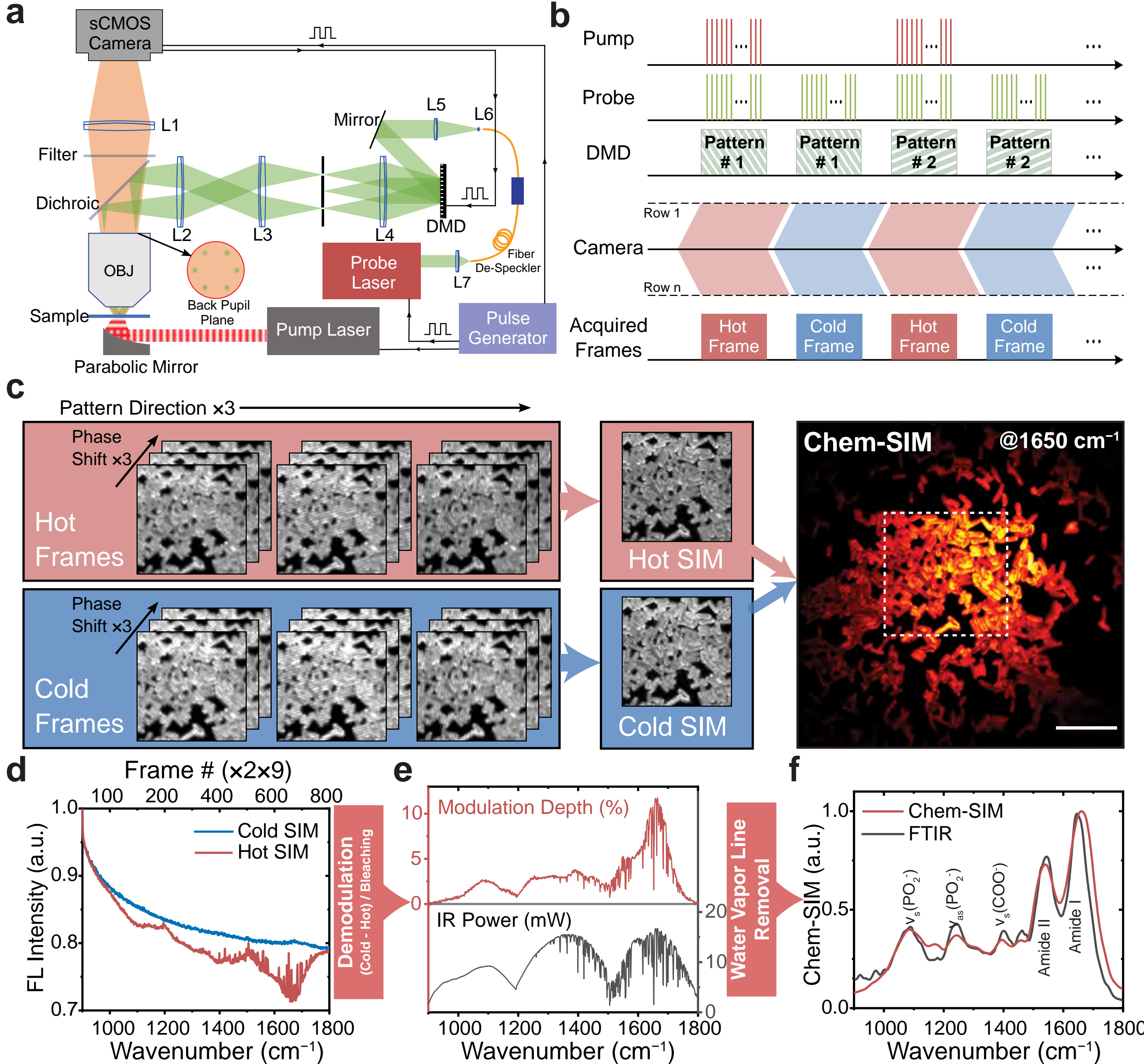

**Figure 1 Principle and workflow of Chem-SIM.** (a) Schematic of the Chem-SIM microscope. (b) Timing diagram for pump, probe, DMD patterns and synchronized camera acquisition. (c) Reconstruction workflow demonstrated using R6G-labelled *E. coli* at $\nu$ = 1650 cm$^{-1}$; scale bar, 10 μm. (d) Mean fluorescence intensity within the dashed boxed region for hot-SIM and cold-SIM hyperspectral stacks. (e) Bleaching-corrected raw fluorescence modulation-depth spectrum and on-sample IR power. (f) Chem-SIM absorption spectrum after water-vapor line removal by weighted least-squares normalization, compared with a normalized reference FTIR spectrum.

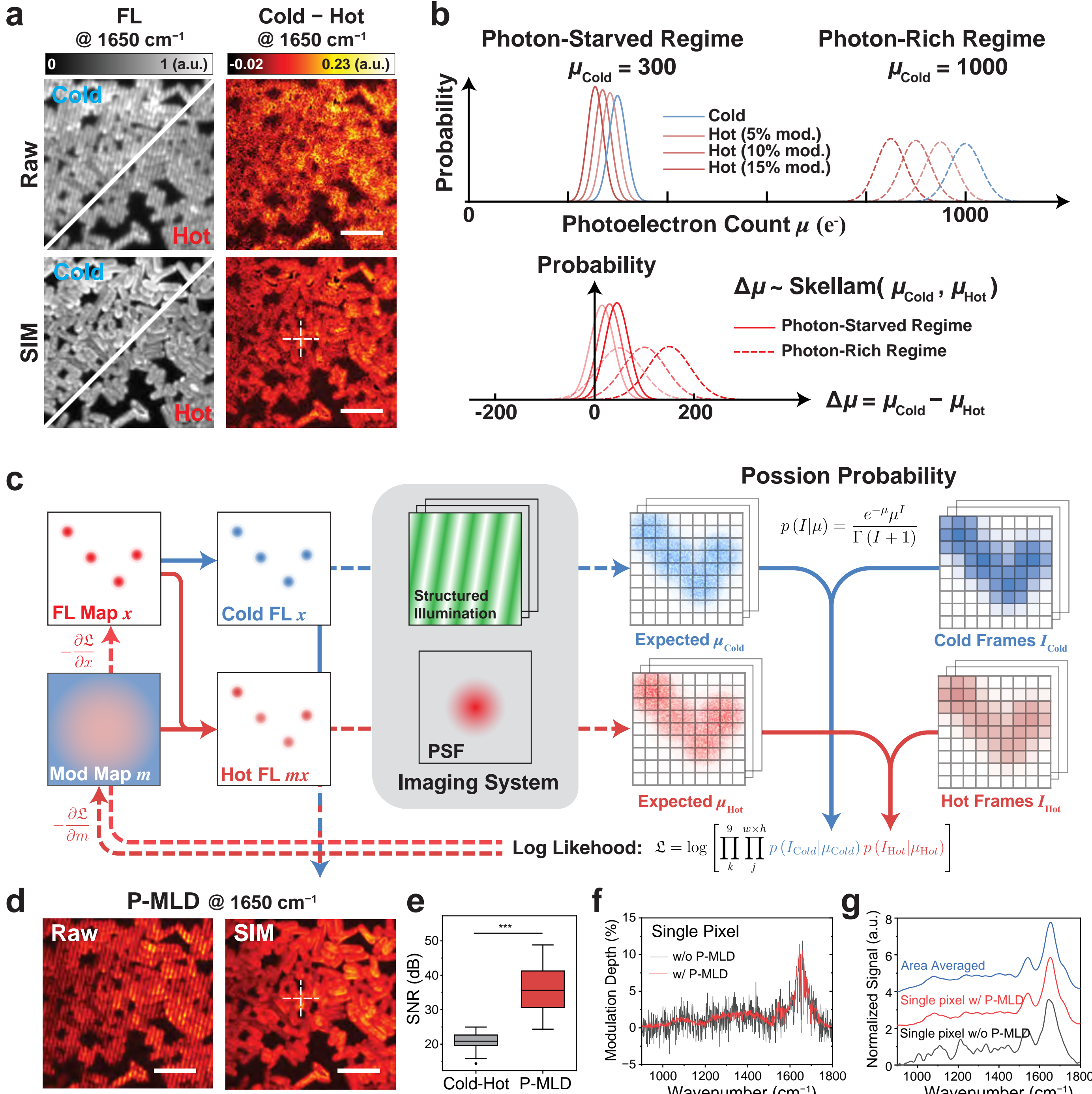

**Figure 2 Poisson maximum-likelihood demodulation (P-MLD) enabled high-quality Chem-SIM.** (a) Cold, Hot, and Cold − Hot images from the dashed boxed region in **Figure 1c**, showing that simple subtraction yields poor SNR in both raw patterned and SIM-reconstructed modulation images. (b) Schematic 1D intensity distributions illustrating loss of modulation contrast when subtracting Poisson-distributed hot and cold photon counts; the pixel-wise difference follows a Skellam distribution that mixes contributions from regions with different fluorescence levels and modulation depths. (c) Image-formation model and P-MLD workflow: a fluorescence map $x$ and fractional modulation map $m$ generate hot fluorescence mx; structured illumination and the microscope PSF produce expected cold and hot photon counts $\mu_{Cold}$ and $\mu_{Hot}$, and $x$ and $m$ are jointly estimated by maximizing the Poisson log-likelihood. (d) P-MLD improves the SNR in both raw patterned and SIM-reconstructed modulation images. (e) SNR comparison between Cold − Hot and P-MLD. (f) Single pixel raw modulation-depth spectrum at the position of dashed cross w/o and w/ P-MLD, $p = 0.00195$. (g) Area-averaged and single-pixel Chem-SIM spectra showing that P-MLD enables high-SNR vibrational fingerprints at the single-pixel level. Scale bars, 5 μm.

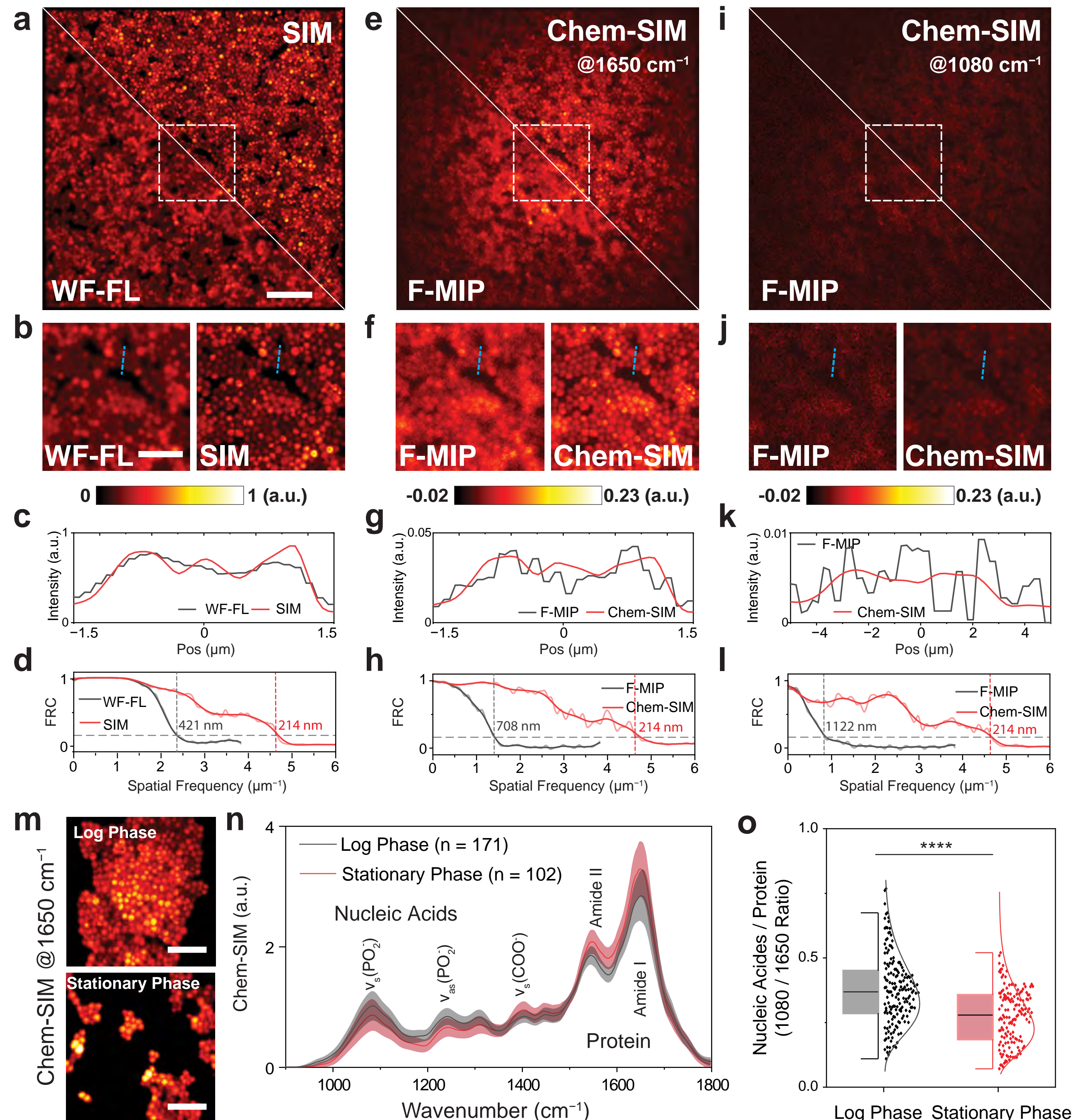

**Figure 3 Spatial and spectroscopic Chem-SIM mapping of *S. aureus*.** (a-d) Wide-field fluorescence (WF-FL) and SIM reconstruction of the same field of view, shown as (a) split-views overlays, (b) zoomed-in crops of the dashed box, (c) line profiles along the blue dashed lines, and (d) Fourier ring correlation (FRC) curves quantifying the resolution gain. (e-h) Corresponding F-MIP (Cold−Hot) and Chem-SIM (with P-MLD) images at the amide-I band ($\nu$ = 1650 $cm^{-1}$), displayed with the same (e) split-view, (f) zoom, (g) line-profile and (h) FRC layout. (i-l) F-MIP and Chem-SIM images at the nucleic-acid band ($\nu$ = 1080 $cm^{-1}$), likewise shown as (i) split views, (j) zooms, (k) line profiles, and (l) FRC analysis. (m-o) Growth-phase analysis: (m) Chem-SIM maps @ 1650 $cm^{-1}$ for stationary-phase and log-phase populations; (n) area-normalized population-averaged single-bacterium Chem-SIM spectra (mean ± s.e.m.) for the two phases; (o) distribution of the 1080/1650 $cm^{-1}$ Chem-SIM intensity ratio (nucleic acids/protein) comparing log-phase and stationary-phase cells, $p$ = 1.94×10−10. Scale bars, 10 µm (a) and 5 µm (b, m).

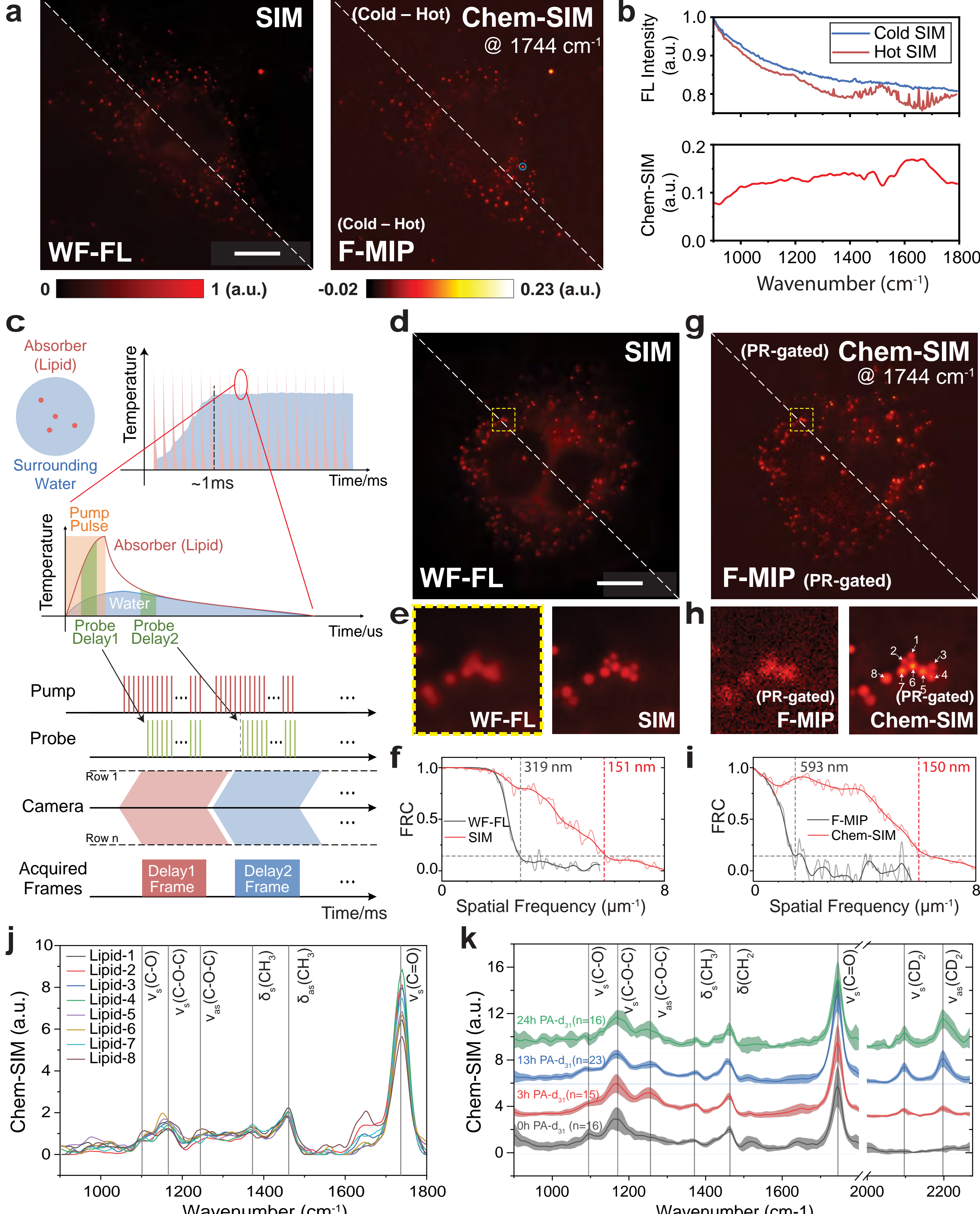

**Figure 4 Background-suppressed Chem-SIM of cells in aqueous environments via photothermal relaxation (PR) gating.** (a) Hela cell imaged by WF-FL and SIM (split view) and corresponding F-MIP and Chem-SIM at 1744 $cm^{-1}$ using the hot-cold mode. (b) Hyperspectral Chem-SIM from the blue-circled lipid droplet in (a), showing Cold-/Hot- SIM intensity traces and corresponding normalized spectrum. (c) Conceptual diagram and timing scheme of PR gating. (d-f) WF-FL/SIM reconstructions for comparison, using frames acquired at delay 2: (d) WF-FL/SIM split view, (e) zoomed-in of the yellow dashed box, and (f) FRC analysis. (g-i) PR-gated F-MIP and Chem-SIM images at 1744 $cm^{-1}$: (g) full-field view, (h) zoomed-in view, and (i) FRC analysis. (j) Area-normalized single-droplet Chem-SIM spectra from the eight droplets in (h), showing consistent ester C=O peaks, strongly suppressed out-of-band background, and additional lipid bands with varying relative amplitudes. (k) Area-normalized Chem-SIM spectra of PA-d31–treated OVCAR5 cells under different incubation conditions. Scale bars, 10 μm.

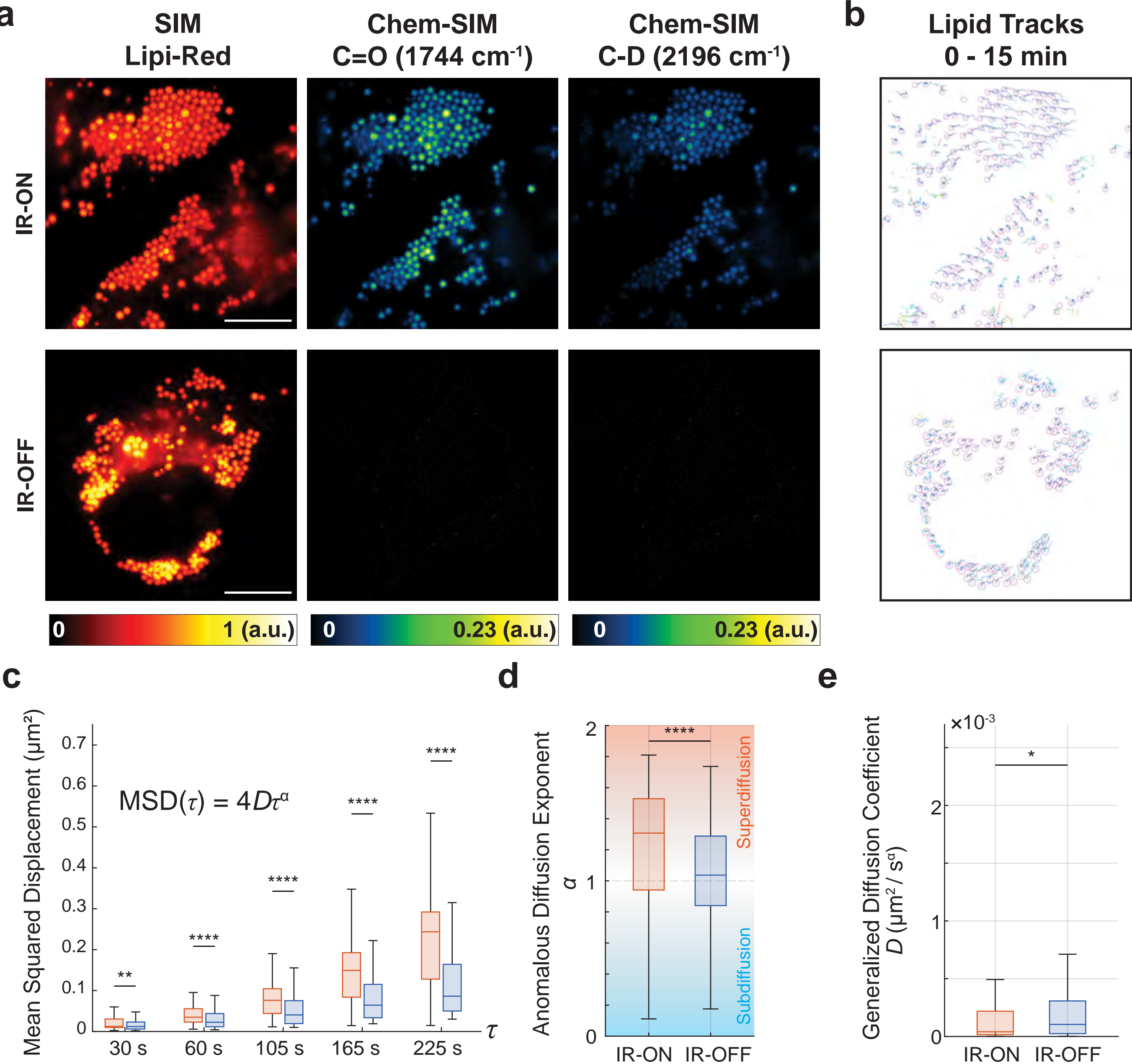

**Figure 5** **Chem-SIM imaging on PA-$d_{31}$–treated live OVCAR5 cells.** (a) SIM Lipi-Red images (left), Chem-SIM map at the ester C=O band (1744 $cm^{-1}$, middle) and the C–D band (2196 $cm^{-1}$, right). (b) Lipid-droplet tracks over 0–15 min for PA-d31–loaded OVCAR5 cells imaged with the IR pump on (IR-ON, top) or off (IR-OFF, bottom). (c) Mean squared displacement (MSD) of individual droplets versus lag time $\tau$, summarized as box plots for IR-ON (orange) and IR-OFF (blue), $p = 5.0\times10^{-3}$, $4.8\times10^{-6}$, $3.1\times10^{-9}$, $6.0\times10^{-13}$, and $6.0\times10^{-13}$ for $\tau = 30$, 60, 105, 165, and 225 s, respectively. (d) Distribution of the anomalous diffusion exponent $\alpha$ for IR-ON and IR-OFF trajectories, with background shading indicating subdiffusion ($\alpha < 1$) and superdiffusion ($\alpha > 1$) regimes, $p = 2.5\times10^{-6}$. (e) Generalized diffusion coefficient $D$ for the same trajectories, showing modest changes between IR-ON and IR-OFF conditions, $p = 0.040$. Error bars and boxes denote the spread across all tracked droplets. Scale bars, 10 μm.

# Supplementary Information

# Chem-SIM: Super-resolution Chemical Imaging of Organelles via Photothermal Modulation of Structured-Illumination Fluorescence

Dashan Dong[1,2], Danchen Jia[1,2], Xinyan Teng[3], Jianpeng Ao[1,2], George Abu-Aqil[1], Biwen Gao[3], Meng Zhang[1,2], Qing Xia[1,2], Ji-Xin Cheng[1,2,3,4,*]

[1] Department of Electrical and Computer Engineering, Boston University, Boston, Massachusetts 02215, United States;

[2] Photonics Center, Boston University, Boston, Massachusetts 02215, United States;

[3] Department of Chemistry, Boston University, Boston, Massachusetts 02215, United States;

[4] Department of Biomedical Engineering, Boston University, Boston, Massachusetts 02215, United States.

* Correspondence: jxcheng@bu.edu (J.-X.C.)

# Supplementary Note 1: SIM Reconstruction preserving Photothermal Modulation

Many algorithms have been proposed for SIM reconstruction, yet suppressing artifacts while preserving quantitative fidelity remains challenging[1]. Classical SIM algorithms typically first form linear differences (solving linear inverse equation) of multiple phase-shifted stripe images to demodulate the illumination pattern and separate the frequency-shifted object bands, and then recombine these components into a super-resolved image. However, these differential operations also amplify shot noise and can distort the temporal intensity profile of the signal. Moreover, to further attenuate out-of-focus background and sharpen resolution, frequency-domain filtering is often adopted, which introduces multiple hyperparameters that require dataset-specific fine-tuning and can alter image statistics[2]. Indeed, many state-of-the-art SIM pipelines—such as generalized Wiener reconstructions[3], apodized OTF weighting[4, 5], and optical-sectioning (OS-SIM) variants[6]—explicitly rely on frequency-domain weighting and notching to suppress noise and out-of-focus leakage.

For Chem-SIM, however, the photothermal-induced fluorescence modulation encodes small, bond-selective changes on top of both structured-illumination signal and background with changing noise, so any differential operation and aggressive filtering that reshapes the frequency spectrum can easily distort this modulation. Consistent with this expectation, we find that conventional Wiener-based inversions often visibly distort this component (**Supplementary Figure 1**). To avoid such filter-induced bias, we therefore adopt a joint Richardson–Lucy (RL) SIM reconstruction[7-9], without any additional spatial filtering, which explicitly models Poisson photon statistics and is insensitive to frame order, thereby preserving the modulation signal to the greatest extent. Practically, this also eliminates the need to hand-tune frequency-domain weights, providing a simple and robust reconstruction pipeline for Chem-SIM.

## Forward model

Let $x(\mathbf{r}) \geq 0$ denote the high-resolution fluorophore distribution. For SIM frame $k = 1, \ldots, K$, a known illumination pattern $s_k(\mathbf{r}) \geq 0$ modulates the object and the microscope PSF $h(\mathbf{r})$ blurs the scene. The mean photon rate on the camera is

$$\mu_k(\mathbf{r}) = (h \otimes [\, s_k \odot x \,])(\mathbf{r}) + b(\mathbf{r}) \tag{S1}$$

where $\otimes$ is convolution, $\odot$ is pointwise product, and $b(\mathbf{r}) \geq 0$ is the background/offset. The measurements $I_k(\mathbf{r}) \in \mathbb{N}$ follow Poisson statistics,

$$I_k(\mathbf{r}) \sim \text{Poisson}\big(\mu_k(\mathbf{r})\big). \tag{S2}$$

For implementation we use the discrete form

$$\boldsymbol{\mu}_k = H\,(S_k\,\boldsymbol{x}) + \boldsymbol{b}_k, \tag{S3}$$

with $H$ the convolution operator (or OTF in Fourier domain), $S_k = \text{diag}(s_k)$, and $x \geq 0$.

## Poisson likelihood and RL objective

The negative log-likelihood (up to constants) over all frames $(k = 1, \dots, K)$ and pixels $(j = 1, \dots, w \times h)$ is

$$\mathcal{L}(x) = \sum_{k=1}^{K} \sum_{j=1}^{w \times h} \left[ \mu_{k,j} - I_{k,j} \log \mu_{k,j} \right], \qquad \mu_{k,j} = H(S_k\, \boldsymbol{x})_j + b_i. \tag{S4}$$

This is the standard Poisson data-fidelity term for photon-counting imaging. Richardson–Lucy (RL) algorithm can be derived by applying an expectation–maximization (EM)–style majorize–minimize scheme to this Poisson objective, which yields a multiplicative, positivity-preserving update for $(\boldsymbol{x})$ that naturally respects the non-negativity of fluorophore densities.

## Joint RL–SIM update

Setting $\partial\mathcal{L}/\partial\boldsymbol{x} = 0$ and applying the standard RL surrogate gives the joint SIM update

$$\boldsymbol{x}^{(t+1)} = \boldsymbol{x}^{(t)} \odot \frac{\sum_k \mathrm{S}_k \left[ H^\top \left( \frac{\boldsymbol{I_k}}{\boldsymbol{\mu_k}^{(t)}} \right) \right]}{\sum_k S_k [H^\top \mathbf{1}]}, \boldsymbol{\mu}_k^{(t)} = H\left(S_k \boldsymbol{x}^{(t)}\right) + b. \tag{S5}$$

Here 1 is an all-ones vector, $H^\top$ is the adjoint convolution, and the division is pointwise. The numerator back-projects the ratio image $\boldsymbol{I}_k/\boldsymbol{\mu}_k^{(t)}$ through the adjoint optics and demodulates by $S_k$; the denominator normalizes by the back-projected DC response. Summation over $k$ makes this a joint reconstruction: all phase/orientation frames drive a single object $\boldsymbol{x}$. In this form, Chem-SIM reconstruction is conceptually identical to standard RL deconvolution, but extended to jointly handle multiple structured-illumination frames while explicitly preserving the small photothermal modulation encoded in each frame.

In practice, our SIM reconstruction further accounts for the mismatch between the high-resolution reconstruction grid and the lower-resolution camera sampling by explicitly modeling the downsampling/upsampling operators in the image-formation process. We also introduce per-frame illumination scale factors $\alpha_k$ to correct for orientation-dependent diffraction efficiency of the DMD patterns, ensuring that differences in stripe direction do not bias the recovered modulation amplitude. In both the high-resolution fluorophore estimate $x$ and the background estimate $b$, we impose non-negativity constraints at each iteration so that the multiplicative RL updates remain physically meaningful and numerically stable, preventing divergence of the iterative scheme. The complete discrete implementation of this joint RL–SIM scheme, including sampling operators, background refinement, illumination scaling, and non-negativity enforcement, is summarized in **Algorithm 1**.

**Algorithm 1:** Joint RL–SIM reconstruction

**Input:** Raw SIM frames $\{I_k\}_{k=1}^K$; illumination patterns $\{s_k\}_{k=1}^K$; PSF $h$ and adjoint $h^\top$; downsampling operator $D_r$ and adjoint $D_r^\top$; max iterations $T_{\max}$, numerical guard $\varepsilon > 0$.

**Output:** High-resolution fluorophore distribution $x^\star \geq 0$.

**Initialization:**

Initialize the object from the widefield average of all frames: $x^{(0)} = D_r^\top\left(\dfrac{1}{K}\sum_{k=1}^K I_k\right)$.

Set illumination scale factors $\alpha_k^{(0)} = 1$.

Initialize the background $b^{(0)} = 0$.

Precompute the backprojected DC response for RL normalization: $c = h^\top * D_r^\top \mathbf{1}$.

**for** $t = 1$ **to** $T_{\max}$ **do**

*(1) Forward model and positivity of the mean:*

$$\mu_k^{(t)} = D_r\big(h * (\alpha_k^{(t)}\, s_k \odot x^{(t)})\big) + b^{(t)}, \quad \text{then } \mu_k^{(t)} \leftarrow \max(\mu_k^{(t)}, \varepsilon).$$

*(2) Ratio images and upsampling:*

$$\rho_k^{(t)} = I_k./\mu_k^{(t)}, \quad \hat{\rho}_k^{(t)} = D_r^\top(\rho_k^{(t)}).$$

*(3) Joint backprojection and RL denominator:*

RL numerator: $N^{(t)} = \displaystyle\sum_{k=1}^K \alpha_k^{(t)}\, s_k \odot \big(h^\top * \hat{\rho}_k^{(t)}\big)$;

RL denominator: $D^{(t)} = \displaystyle\sum_{k=1}^K \alpha_k^{(t)}\, s_k \odot c$;

Enforce positivity: $D^{(t)} \leftarrow \max(D^{(t)}, \varepsilon)$.

*(4) RL multiplicative update of the object:*

$$x^{(t+1)} = x^{(t)} \odot \frac{N^{(t)}}{D^{(t)}}, \quad x^{(t+1)} \leftarrow \max(x^{(t+1)}, 0).$$

*(5) Background update:*

$$b^{(t+1)} = h * \left(\frac{1}{K}\sum_{k=1}^K \left[I_k - D_r\big(h * (\alpha_k^{(t)}\, s_k \odot x^{(t+1)})\big)\right]\right),$$

Enforce positivity: $b^{(t+1)} \leftarrow \max(b^{(t+1)}, 0)$.

*(6) Per-frame illumination factor update:*

Predicted frame with unit scale: $\hat{\mu}_k^{(t+1)} = D_r\big(h * (s_k \odot x^{(t+1)})\big)$;

Update illumination scale factor: $\alpha_k^{(t+1)} = \dfrac{I_k - b^{(t+1)}}{\hat{\mu}_k^{(t+1)} + \varepsilon}$.

**return** $x^\star = x^{(T_{\max})}$.

## Supplementary Note 2: Imaging Conditions

During hyperspectral mid-IR wavenumber scans, the imaging rate is limited by the QCL's wavelength-switching time; each change introduces a ~200 ms of dead time during which neither the IR pump nor the probe illuminates the sample. Once the IR pump is set to a given wavenumber, the frame rate is primarily capped by the camera and the chosen imaging region of interest (ROI). We operate the camera and DMD in a hardware-synchronized triggering mode to minimize dead time from rolling-shutter readout, which also scales with the number of lines in the imaging ROI. Owing to camera readout and DMD trigger latencies, the effective exposure per frame is governed by the DMD on-time rather than the nominal camera exposure. For **Figures 1–3** we use 40 fps at a 512×512-pixel ROI, yielding an effective exposure of 23.1 ms per frame (corresponding to 923 pump pulses). For **Figures 4–5** and the **Supplementary Movie 2** we use 10 fps at a 768×768-pixel ROI, with an effective exposure of 90.5 ms per frame (4505 pulses). Finally, to suppress timing jitter, we insert a small guard interval so that each exposure integrates an identical, stable integer number of pump pulses. The imaging conditions is detailed in **Table 1**.

**Table 1 Imaging Conditions for All Figures**

| Figure | IR Wavenumber Range/Interval ($cm^{-1}$) | Camera Frame Rate (fps; ROI) | Effective Exposure Time (ms; # pump pulses) | Total Frames (#) | Total Acquisition Time (s) |
|---|---|---|---|---|---|
| Figure 1-3 | 900-1798 / 1 | 40 (512×512) | 23.1 (923 pulses) | 2×9×899 | 585 |
| Figure 4a-j | 900-1792 / 4 | 10 (768×768) | 90.5 (4505 pulses) | 2×9×224 | 448 |
| Figure 4k | 900-1792 / 4<br>2000-2268 / 4 | 10 (768×768) | 90.5 (4505 pulses) | 2×9×(224+68) | 584 |
| Figure 5a & Supplementary Movie 2 | 1744 & 2196 | 10 (768×768) | 90.5 (4505 pulses) | 2×9×2×60 | 3.8* |

*For Figure 5a and the Supplementary Movie 2, 3.8 s is the acquisition time per time point (36 frames per time point at 10 fps). The full time-lapse dataset comprises 60 time points with 15 s interval (15 min in total).

## Supplementary Note 3: Weighted Least Squares Spectral Normalization Algorithm

To compensate mid-IR power variation (including water-vapor lines) and remove baseline offsets while preserving sharp spectral features in 1D MIP/Chem-SIM spectra. We solve a weighted least squares spectral normalization problem, detailed as follow:

Given a 1D spectrum $y \in \mathbb{R}^N$ measured on wavenumber grid $\nu_i$ and the on-sample mid-IR power profile $p_i$, we normalize the power as

$$w_i = \frac{p_i}{\max_j p_j} \in [0,1], \tag{S6}$$

and estimate a nonnegative normalized spectrum $u \geq 0$ and a scalar baseline $b$ by

$$\min_{u \geq 0,\, b} \Phi(u,b) = \sum_{i=1}^{N} (w_i u_i - y_i - b)^2 + \lambda \mathrm{TV}(u) + \mu |b|. \tag{S7}$$

Where $\mathrm{TV}(u) = \sum_{i=1}^{N-1} |(u_{i+1} - u_i)/\Delta\nu_i|$ is the 1D total-variation seminorm on the non-uniform grid $\Delta\nu_i = |\nu_{i+1} - \nu_i|$. If regularization and baseline are ignored, (S7) reduces to the element-wise stabilized division $u \approx \frac{y}{w}$; the TV term preserves sharp, edge-preserving spectra and $|b|$ removes a constant offset.

Optimization of (S7) is performed with a fast iterative shrinkage–thresholding (FISTA) scheme[10] (**Algorithm 2)**, using a primal–dual hybrid gradient (PDHG) proximal[11] (**Algorithm 3**) for the TV term and soft-thresholding for the baseline. In this work we use $\eta = 0.5, \lambda = 0.05, \mu = 10^{-3}$, with $K = 50$ outer FISTA iterations and $T = 20$ inner TV iterations. Within a reasonable parameter range, the output is stable and shows no significant differences.

We impose homogeneous Neumann boundary conditions (zero normal derivative at the two ends of the 1D wavenumber axis) for the TV operator. This keeps endpoints free, regularizes interior jumps without edge bias, and avoids injecting artificial trends at the boundaries. Accordingly (Algorithm 2), we use forward difference $\nabla\colon \mathbb{R}^N \to \mathbb{R}^{N-1}$ and a divergence $div : \mathbb{R}^{N-1} \to \mathbb{R}^N$ that is the negative adjoint of $\nabla$ under standard Euclidean inner product, consistent with Neumann boundary conditions:

$$(\nabla \mathrm{u})_i = \frac{u_{i+1} - u_i}{\Delta\nu_i}, \quad i = 1, \dots, N-1, \tag{S8}$$

$$(\operatorname{div} p)_1 = -\frac{p_1}{\Delta\nu_1}, \quad (\operatorname{div} p)_i = -\frac{p_i}{\Delta\nu_i} + \frac{p_{i-1}}{\Delta\nu_{i-1}} \quad (2 \leq i \leq N-1), \tag{S9}$$

These operators satisfy $\langle \nabla u, p \rangle = -\langle u, \operatorname{div} p \rangle$ and implement the usual "reflect (mirror) padding" interpretation of Neumann boundary conditions in 1D.

**Algorithm 2:** FISTA solver for Weighted Least Squares Spectral Normalization

**Input:** Raw spectrum $y \in \mathbb{R}^N$, weights $w \in [0,1]^N$, wavenumber grid $\{\nu_i\}$; step size $\eta$, TV weight $\lambda$, baseline weight $\mu$; outer iters $K$, inner TV iters $T$.

**Output:** Nonnegative normalized spectrum $u$, baseline $b$.

Normalize power: $w \leftarrow w / \max(w)$.

Compute spacings $\Delta\nu_i = |\nu_{i+1} - \nu_i|$.

Initialize $b \leftarrow 0$, $x \leftarrow y - b$, $u \leftarrow x$, $t \leftarrow 1$.

**for** $k = 1, \ldots, K$ **do**

$t_{\text{new}} \leftarrow \dfrac{1 + \sqrt{1 + 4t^2}}{2}$.

$g \leftarrow w \odot (w \odot u - y - b)$.

$f \leftarrow u - \eta\, g$.

$u_{\text{new}} \leftarrow \text{Prox_TV}(f, \lambda, T, \Delta\nu)$.

$u \leftarrow u_{\text{new}} + \dfrac{t - 1}{t_{\text{new}}} (u_{\text{new}} - x)$.

$x \leftarrow u_{\text{new}},\ t \leftarrow t_{\text{new}}$.

$r \leftarrow \text{mean}\big(w \odot u - y\big)$.

$b \leftarrow \text{sign}(r)\ \max(|r| - \mu, 0)$.

**return** $u, b$.

**Algorithm 3:** Prox_TV: Proximal for $\min_u \frac{1}{2}\|u - f\|^2 + \lambda\, \text{TV}(u)$

**Input:** $f \in \mathbb{R}^N$, $\lambda > 0$, inner iterations $T$, spacings $\Delta\nu$.

**Output:** Denoised, nonnegative $u$.

Initialize $u \leftarrow f$, $p \leftarrow 0 \in \mathbb{R}^{N-1}$, $p_{\text{prev}} \leftarrow p$, $\theta \leftarrow 1$.

Set Lipschitz bound $L \leftarrow 4$.

**for** $t = 1, \ldots, T$ **do**

$\tilde{p} \leftarrow p + \dfrac{1}{L\,\lambda} \nabla\big(u - \lambda\ \text{div}(p)\big)$.

$p_{\text{new}} \leftarrow \text{clip}(\tilde{p}, -1, 1)$.

$\theta_{\text{new}} \leftarrow \dfrac{1 + \sqrt{1 + 4\theta^2}}{2}$.

$p \leftarrow \left(1 + \dfrac{\theta - 1}{\theta_{\text{new}}}\right) p_{\text{new}} - \dfrac{\theta - 1}{\theta_{\text{new}}}\, p_{\text{prev}}$.

$p_{\text{prev}} \leftarrow p_{\text{new}},\ \theta \leftarrow \theta_{\text{new}}$.

$u \leftarrow f - \lambda\ \text{div}(p)$.

$u \leftarrow \max(u, 0)$.

**return** $u$.

## Supplementary Note 4: Poisson maximum-likelihood demodulation (P-MLD)

With the Cold-state image corresponds to the baseline fluorescence map $x_{\text{Cold}} = x$, and the Hot fluorescence map is given by $x_{\text{Hot}} = m \odot x$. P-MLD jointly estimates modulated image $z \equiv x - m \odot x,$ by maximizing the joint Poisson log-likelihood[9]:

$$\mathcal{L}(x,m) = \log\left[\prod_{k}^{9}\prod_{j}^{w\times h} p\left(I_k^{\text{Cold}}|\mu_k^{\text{Cold}}\right) p\left(I_k^{\text{Hot}}|\mu_k^{\text{Hot}}\right)\right]. \tag{S10}$$

Here, $j$ indexes pixels, $w \times h$ is the image size, and $I_k^{\text{Cold}}, I_k^{\text{Hot}}$ are observed images acquired from expected photon counts

$$\mu_k^{\text{Cold}} = A_k(x),\ \ \mu_k^{\text{Hot}} = A_k(m\odot x) \tag{S11}$$

for the Cold/Hot frames. In this formulation, ***we impose a strong prior that the Cold and Hot measurements arise from the same underlying fluorescence distribution $x$***, and introduce the modulation map $m$ multiplicatively rather than through explicit differencing; this tight coupling between the two channels is crucial for enhancing the SNR of the recovered modulation image.

Following **Supplementary Note 1**, the SIM microscope is modeled by the linear operator

$$A_k(x) = h * (s_k \odot x) + b, \tag{S12}$$

where $h$ is the PSF, $s_k$ is the $k$-th illumination pattern, $b$ is a slowly varying background, and $*$ denotes convolution.

By inserting the Poisson probability mass function

$$p(I|\mu) = \frac{e^{-\mu}\mu^I}{\Gamma(I+1)}, \tag{S12}$$

into (S10), the product of probabilities can be rewritten as a sum of log-likelihood terms:

$$\mathcal{L}(x,m) \propto \sum_k\sum_j\left[I_{k,j}^{\text{Cold}}\log\mu_k^{\text{Cold}} - \mu_k^{\text{Cold}} + I_{k,j}^{\text{Hot}}\log\mu_k^{\text{Hot}} - \mu_k^{\text{Hot}}\right], \tag{S13}$$

P-MLD maximize $\mathcal{L}$ (equivalently, minimize the negative log-likelihood) under nonnegativity constraints. To improve robustness under low photon budgets and to stabilize high-frequency content, we add mild regularization, and jointly optimize the background $b$:

$$\min_{x\geq 0, 0\leq m\leq 1, b\geq 0} \ell(x,m,b) = -\mathcal{L}(x,m,b)\ +\ \lambda_m R_m(xm) + \lambda_x R_x(x), \tag{S14}$$

where spatial continuity is promoted with a 2D Hessian penalty on $x$ and a 2D TV penalty on the modulation map $m$.

The problem is solved with a nonnegative Richardson–Lucy (RL)-style multiplicative scheme[12], alternating updates for $x$, $m$, and $b$, as detailed in **Algorithm 4**. The final Chem-SIM modulation image is then obtained as $z \equiv x - m \odot x$.

P-MLD can also demodulate wide-field fluorescence images by assuming uniform illumination (i.e., $s_k = 1$), yielding the "Raw" result in **Figure 2d**. In the actual implementation, P-MLD uses the same downsampling/upsampling operators and illumination scaling corrections as described in Supplementary Note 1; for brevity, these details are not repeated in the pseudocode.

**Algorithm 4:** Poisson maximum-likelihood demodulation (P-MLD) with SIM

**Input:** Cold / Hot Raw frames $\{I_k^{\text{Cold}}, I_k^{\text{Hot}}\}_{k=1}^{K}$; illumination patterns $\{s_k\}_{k=1}^{K}$; PSF $h$ and adjoint $h^\top$; max iterations $T_{\text{max}}$; small numerical guard $\varepsilon > 0$.

**Output:** Original fluorescence image $x^\star \geq 0$, Photothermal image $z^\star = x^\star - m^\star \odot x^\star$.

**Initialization:**

Initialize $x$ with the average of Cold stack: $x^{(0)} = \text{mean}(\{I_k^{\text{Cold}}\})$.

Initialize the modulation map and enforce bounds: $m^{(0)} = \mathbf{1}$.

Initialize the background: $b^{(0)} = 0$.

Define the SIM forward operator and its adjoint:
$A_k(\cdot) = h * (s_k \odot \cdot)$, $A_k^\top$ is its adjoint.

**for** $t = 1$ **to** $T_{\text{max}}$ **do**

*(1) Forward model for Cold / Hot means:*

$$\mu_k^{\text{Cold},(t)} = A_k(x^{(t)}) + b^{(t)},$$
$$\mu_k^{\text{Hot},(t)} = A_k\big(m^{(t)} \odot x^{(t)}\big) + b^{(t)}.$$

Enforce positivity: $\mu_k^{\text{Cold},(t)} \leftarrow \max(\mu_k^{\text{Cold},(t)}, \varepsilon)$, $\mu_k^{\text{Hot},(t)} \leftarrow \max(\mu_k^{\text{Hot},(t)}, \varepsilon)$.

*(2) Ratio images:*

$$r_k^{\text{Cold}} = I_k^{\text{Cold}}./\mu_k^{\text{Cold},(t)}, \quad r_k^{\text{Hot}} = I_k^{\text{Hot}}./\mu_k^{\text{Hot},(t)}.$$

*(3) RL-style multiplicative updates for $x$ and $m$:*

$$x^{(t+1)} = x^{(t)} \odot \frac{\displaystyle\sum_{k=1}^{K} s_k \odot A_k^\top\big(r_k^{\text{Cold}} + m^{(t)} \odot r_k^{\text{Hot}}\big)}{\displaystyle\sum_{k=1}^{K} s_k \odot A_k^\top\big(\mathbf{1} + m^{(t)}\big)}, \quad x^{(t+1)} \leftarrow \max(x^{(t+1)}, \varepsilon).$$

$$m^{(t+1)} = m^{(t)} \odot \frac{\displaystyle\sum_{k=1}^{K} s_k \odot A_k^\top\big(r_k^{\text{Hot}} \odot x^{(t)}\big)}{\displaystyle\sum_{k=1}^{K} s_k \odot A_k^\top\big(x^{(t)}\big)}, \quad m^{(t+1)} \leftarrow \min\big(\max(m^{(t+1)}, \varepsilon), 1\big).$$

*(4) Denoising / regularization step:*

$$x^{(t+1)} \leftarrow \text{argmin}_{u \geq 0}\Big\{\tfrac{1}{2}\,\|u - x^{(t+1)}\|_2^2 + \lambda_x\, R_x(u)\Big\}.$$

$$m^{(t+1)} \leftarrow \text{argmin}_{0 \leq v \leq 1}\Big\{\tfrac{1}{2}\,\|v - m^{(t+1)}\|_2^2 + \lambda_m\, R_m(v)\Big\}.$$

*(5) Background update:*

$$b^{(t+1)} = h * \left( b^{(t)} \odot \frac{\displaystyle\sum_{k=1}^{K} A_k^\top(r_k^{\text{Cold}} + r_k^{\text{Hot}})}{\displaystyle\sum_{k=1}^{K} A_k^\top(\mathbf{1} + m^{(t)}) + \varepsilon} \right),$$

$$b^{(t+1)} \leftarrow \max(b^{(t+1)}, \varepsilon).$$

**return** $x^\star = x^{(T)}$, $b^\star = b^{(T)}$*, and modulation image* $z^\star = x^\star - m^\star \odot x^\star$.

## Supplementary Note 5: Signal-to-Noise Ratio (SNR) Calculation

We quantify image quality using an ROI-based signal-to-noise ratio (SNR). For each field, we delineated signal ROIs on structures of interest and background ROIs in nearby signal-free regions within the same frame (non-overlapping and comparable in size). Let $\mu_S$ and $\mu_B$ be the mean intensities in the signal and background ROIs, and let $\sigma_B$ be the standard deviation within the background ROI (our noise estimate). We define

$$\mathrm{SNR} = \frac{\mu_S - \mu_B}{\sigma_B}, \quad \mathrm{SNR_{dB}} = 20\log_{10}(\mathrm{SNR}). \tag{S15}$$

retaining the algebraic sign of $\mu$ when reported in linear units. This definition measures contrast normalized by the local background noise and is insensitive to absolute intensity scaling. For group summaries, per-ROI values were aggregated as median ± IQR, and paired ROIs were compared between Cold−Hot and P-MLD SIM images.

## Supplementary Note 6: Fourier ring correlation (FRC) analysis

We quantified spatial resolution using Fourier ring correlation (FRC) computed between independent repeated datasets (WF-FL, SIM, F-MIP, or Chem-SIM). The procedure follows the standard FRC formalism, with a small Wiener-type regularization term added to stabilize the denominator at high spatial frequencies.

Let $I_1(\mathbf{r})$ and $I_2(\mathbf{r})$ be two independently measure of the same field of view, and let

$$\hat{I}_1(\mathbf{q}) = \mathcal{F}\{I_1(\mathbf{r})\}, \quad \hat{I}_2(\mathbf{q}) = \mathcal{F}\{I_2(\mathbf{r})\} \tag{S16}$$

denote their 2D Fourier transforms, with spatial-frequency coordinate $\mathbf{q} = (q_x, q_y)$. We first form the cross-power and power spectra:

$$N(\mathbf{q}) = \Re\{\hat{I}_1(\mathbf{q})\hat{I}_2^*(\mathbf{q})\}, \quad S_1(\mathbf{q}) = \left|\hat{I}_1(\mathbf{q})\right|^2, \quad S_2(\mathbf{q}) = \left|\hat{I}_2(\mathbf{q})\right|^2 \tag{S17}$$

To compute a FRC map, we define

$$\mathrm{FRC}_{\mathrm{grid}}(\mathbf{q}) = \frac{N(\mathbf{q})}{\sqrt{S_1(\mathbf{q})S_2(\mathbf{q})} + \varepsilon}, \tag{S18}$$

where $\varepsilon$ is a small Wiener-like constant to prevents numerical instabilities when both $S_1(\mathbf{q})$ and $S_2(\mathbf{q})$ are dominated by noise at high spatial frequencies and $\sqrt{S_1(\mathbf{q})S_2(\mathbf{q})} \to 0$. In practice, $\varepsilon$ suppresses spurious FRC spikes in the weak-signal regime and yields smoother, more robust FRC curves.

To further reduce pixel-wise fluctuations in Fourier space, $\mathrm{FRC}_{\mathrm{grid}}(\mathbf{q})$ is smoothed with a 2D Gaussian kernel ($\sigma$ = 3 pixels). The final FRC curve is then obtained by radial averaging over concentric rings of constant spatial frequency magnitude $\rho = \|\mathbf{q}\|$:

$$\mathrm{FRC}(\rho) = \langle \mathrm{FRC}_{\mathrm{grid}}(\mathbf{q}) \rangle_{\|\mathbf{q}\|=\rho}, \tag{S19}$$

yielding $\mathrm{FRC}(\rho)$ as a function of spatial frequency $\rho$.

Apart from the inclusion of $\varepsilon$ and Gaussian smoothing, the computation follows the conventional FRC definition, and the reported resolution values correspond to the spatial frequency $\rho$ at which $\mathrm{FRC}(\rho)$ crosses the standard 1/7 threshold.

## Supplementary Note 7: Validation of Transient Photothermal Relaxation in water

### Hot/Cold F-MIP characterization of photothermal relaxation

We quantified photothermal relaxation dynamics at the lipid carbonyl (C=O) band versus water absorption bands using hot/cold fluorescence-detected mid-infrared photothermal (F-MIP) microscope. For hot frames, the pump–probe delay was scanned across the IR pump–probe pulse trains; for cold frames, the IR pump was disabled. To ensure identical probe exposure and readout timing between conditions, only the relative timing of the IR pump was varied. To improve measurement reliability and signal-to-noise while avoiding reconstruction bias, we used 5× averaged F-MIP acquisitions rather than Chem-SIM reconstructions.

Applied to Lipi-Red–stained lipid droplets in HeLa cells, this protocol revealed distinct transient thermal dynamics (normalized by baseline level) at the lipid C=O absorption band (~1744 $cm^{-1}$) and the water band (~1650 $cm^{-1}$) in **Supplementary Figure 3**. The ~1744 $cm^{-1}$ trace reflects the superposition of immediate C=O photothermal absorption and secondary heating from nearby water that has absorbed IR energy, whereas the ~1650 $cm^{-1}$ trace predominantly reports conductive heat transfer from the warmed aqueous medium into the droplet (i.e., no direct lipid absorption). Both traces exhibit an elevated baseline, attributable to slow ambient drift and the build-up of a three-dimensional thermal reservoir in the medium due to gradual spatial heat diffusion.

### Numerical Simulation of Thermal Dynamics

To interpret these observations under photothermal relaxation (PR) gating, we used COMSOL Multiphysics (v6.2) to evaluate transient and cumulative temperature rise in water. A 2D finite-element model comprised a 100-nm-diameter PMMA bead at the $CaF_2$–water interface within a 60 μm (width) × 30 μm (height) domain: a 20 μm water layer atop a 10 μm $CaF_2$ substrate. Perfectly matched layers (PMLs) were applied at the outer electromagnetic boundaries to suppress spurious reflections. The IR pump field was computed with the Electromagnetic Waves interface at 5.787 μm (near the lipid C=O band) and 6.061 μm (near the water band). The beam was modeled as a Gaussian with a $1/e^2$ waist of 30 μm. The resulting absorbed power density served as a volumetric heat source in the Heat Transfer module. Isothermal (room-temperature) conditions were imposed at the outer thermal boundaries to emulate a bulk water bath. Optical (complex refractive index) and thermal properties (thermal conductivity and isobaric heat capacity) for water, PMMA, and $CaF_2$ were taken from COMSOL's material library and public database. The laser excitation was modeled as a nanosecond pulse train (50 kHz repetition rate, 500 ns pulse width). Transient simulations were run for 1.5 ms with a 10 ns time step until a steady-periodic state was reached. The steady-periodic temperature rise was extracted at the bead surface and in the adjacent water, yielding the temperature maps shown in **Supplementary Figure 4 & Supplementary Movie 1**.

**Note.** A 2D spatial simulation model was adopted to reduce computational cost, given the 1.5 ms simulation window required to capture mesoscopic heat diffusion with 10 ns temporal resolution under transient resolution. Because heat flow is restricted to 2D, this model tends to overestimate the local temperature rise compared with the full 3D case. Conversely, the finite simulation domain

and the use of isothermal boundary conditions at its edges can underestimate the peak temperature. As a result, the simulated temperature values should be interpreted only qualitatively, as an illustration of the spatiotemporal evolution of the thermal field. Quantitative estimates of the actual temperature rise can be obtained from fluorescence thermometry, as reported in **Supplementary Figure 10**.

## Supplementary Note 8: Determination of the Photothermal Relaxation (PR) window

**Measurement strategy.** We determined the optimal PR window—defined as the relative timing offset between two pump–probe delay status—using 5× averaged F-MIP imaging of Lipi-Red labelled lipid droplets. Because the readout is purely fluorescence based, the F-MIP signal reports the temperature difference of the droplet between the two delays. In practice, we fixed the pump–probe delay for the first frame (delay-1) and scanned the delay for the second frame (delay-2), yielding delay-dependent transients (**Supplementary Figure 5**).

**Transient dynamics at the lipid band.** At the lipid C=O resonance, all droplets exhibited similarly shaped F-MIP transients. As the delay difference increased, the signal reflected rapid cooling of the droplet toward the surrounding aqueous temperature on the microsecond timescale. After an initial growth and saturation, the signal decreased with further delay, consistent with axial (z-direction) heat transport predicted by simulation. This behavior arises from the combination of shallow IR penetration in water (exponential intensity decay along z) and the large water heat capacity, which together establish a strong axial temperature gradient; axial conduction therefore outpaces lateral (Gaussian) spreading and drives heat redistribution over tens of microseconds.

**Corroboration at the water band and PR window selection.** The water-band transients further substantiate this picture. Across droplets, the F-MIP signal (temperature difference) first becomes negative—indicating that the environment is warmer at delay-2 than at delay-1 and is transferring heat into the droplet—then crosses zero at ~7–12 µs depend on the size of lipid droplets. We take this zero crossing as the optimal PR delay. At longer delays, continued axial heating decay in aqueous medium lower the droplet temperature at delay-2 further, producing a positive F-MIP signal, before both equilibrate thermally.

**Spatial propagation in larger droplets.** For larger lipid droplets, we additionally observed heat flow from the periphery toward the center, consistent with diffusive in-droplet conduction superimposed on the externally driven axial gradient.

**Spectral consequences and practical setting.** As shown in **Supplementary Figure 6**, the PR window strongly modulates the spectral sign and amplitude of the water absorption near ~1650 $cm^{-1}$: shorter PR window yield a negative water peak, whereas increasing the PR window flips the peak to positive contrast. In our experiments, we used an 8 µs PR window, which minimized residual water-background contributions for ~1 µm-diameter droplets.

## Supplementary Note 9: Temperature calibration in cell-culture medium

**Calibration.** To experimentally validate the temperature in the aqueous cellular environment under our mid-IR heating conditions, we implemented fluorescence thermometry on Lipi-Red–stained fixed OVCAR5 cells. As shown in **Supplementary Figure 9**, under standard epifluorescence microscopy the temperature sensitivity of Lipi-Red intensity on cells was calibrated to 1.64±0.03 % $K^{-1}$ using a temperature-controlled stage and a precision fiber thermometer immersed in the culture medium.

**In situ measurement under Chem-SIM conditions.** We then imaged fixed Lipi-Red–stained OVCAR5 cells on the Chem-SIM platform to quantify both the cycle-averaged temperature and the transient thermal dynamics during IR heating. Modulation depths were measured at a 50 kHz IR repetition rate with a fixed 500 ns pulse width in both Cold–Hot mode and PR-gated mode, as summarized in **Supplementary Figure 10**.

In Cold–Hot mode, the additional decrease in fluorescence intensity in the Hot frame, relative to the bleaching baseline, was converted to absolute temperature using the 1.64% $°C^{-1}$ calibration at a laboratory ambient temperature of 20 °C. The resulting baseline (cycle-averaged) temperature at each wavenumber follows the on-sample IR power spectrum. Lipi-Red thermometry reports a maximum temperature of ~31 °C on the lipid droplets during the hyper spectral scanning (**Supplementary Figure 10b**). At 1744 $cm^{-1}$, corresponding to the lipid C=O resonance, the transient thermal dynamics were similarly converted to absolute temperature, revealing distinct heating and cooling kinetics on lipid droplets (**Supplementary Figure 10c**).

In PR-gated mode, both delay-1 and delay-2 sample additional transient temperature rises on top of a baseline similar to that of the Hot frame, such that their fluorescence traces exhibit IR-power-spectrum–shaped modulation superimposed on exponential bleaching (**Supplementary Figure 10d**). By fitting and removing the bleaching component, we obtained the absolute temperatures at the two delays (**Supplementary Figure 10e**). Owing to the double IR dose per cycle in PR-gated mode, the temperature modulation is nearly doubled compared with Cold–Hot mode, reaching a maximum temperature of ~43 °C near the water absorption peak. At the lipid C=O resonance (1744 $cm^{-1}$), the transient PR dynamics yield a peak temperature of 38.4 °C and a cycle-averaged temperature of 33.9 °C (**Supplementary Figure 10f**).

**Supplementary Movie 1 | COMSOL simulation of fast interpulse spatial thermal dynamics.** Simulated temperature distribution and temporal evolution around a 100 nm PMMA bead in water following IR absorption at two characteristic wavelengths: (left) 1650 cm$^{-1}$ corresponding to water absorption, and (right) 1736 cm$^{-1}$ corresponding to the C=O stretch of PMMA.

**Supplementary Movie 2 | Timelapse Chem-SIM imaging on PA-$d_{31}$–treated live OVCAR5 cells.** SIM Lipi-Red images, Chem-SIM map at the ester C=O band (1744 cm$^{-1}$) and the C–D band (2196 cm$^{-1}$), over 0–15 min for PA-d31–loaded OVCAR5 cells imaged with the IR pump on (IR-ON, top) or off (IR-OFF, bottom).

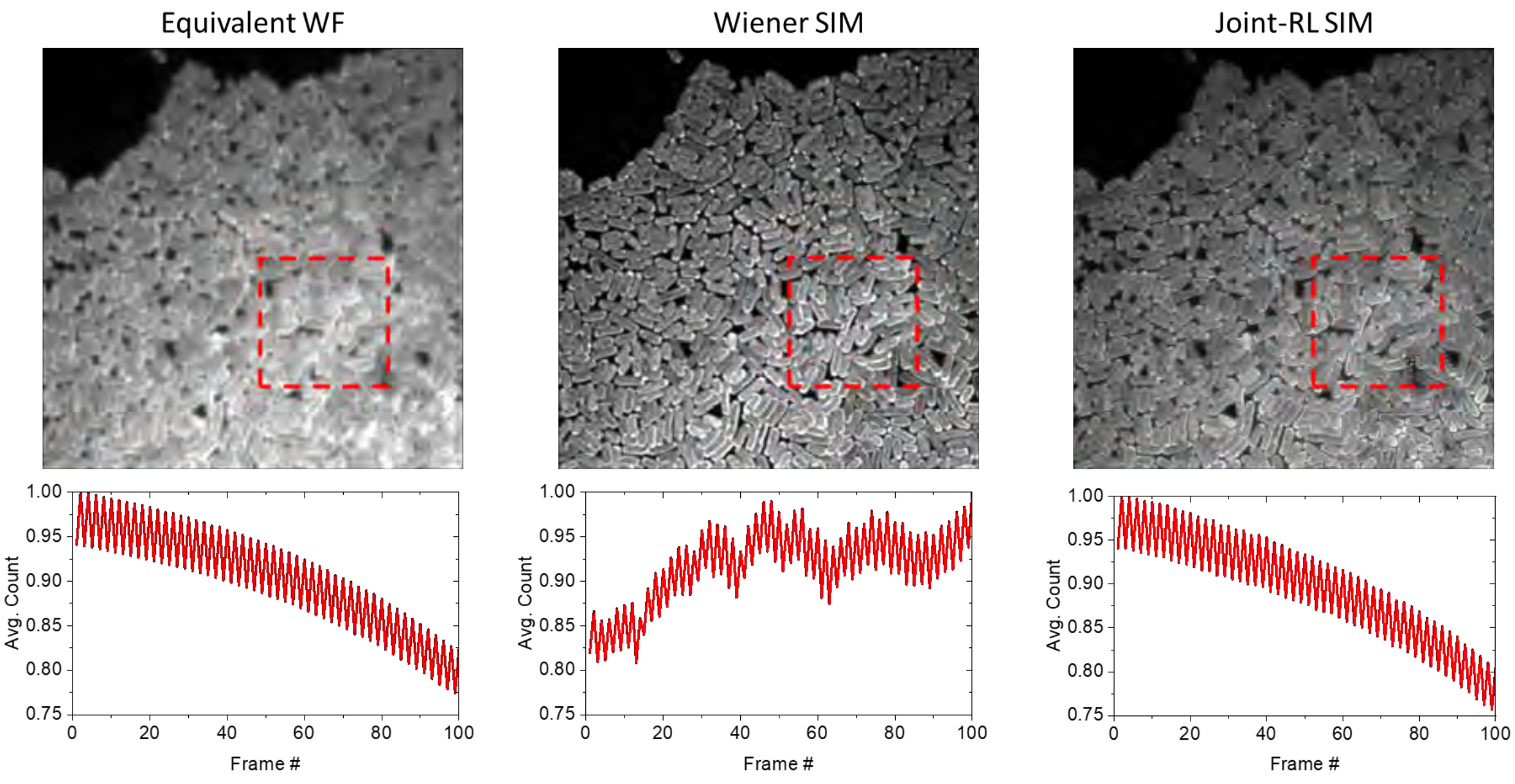

**<span style="color:#1f6fbf">Supplementary Figure 1</span> | Comparison of modulation-signal recovery across SIM reconstruction methods.**

(Top) Reconstructed images from Equivalent wide-field (WF), Wiener-SIM, and Joint RL–SIM. (Bottom) Mean fluorescence-intensity profiles extracted from the region enclosed by the red dashed box, highlighting differences in modulation preservation.

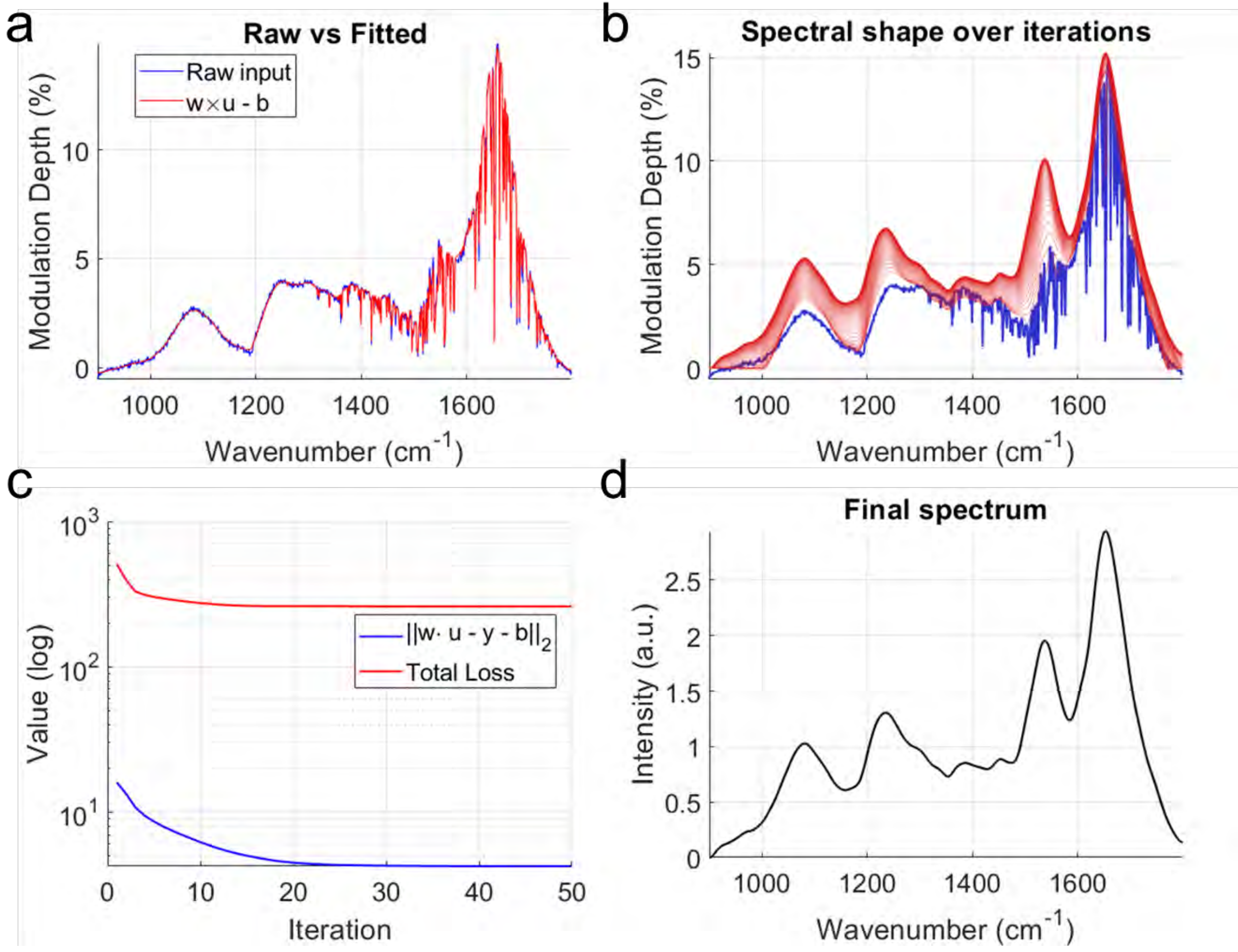

**Supplementary Figure 2 | Weighted least squares spectral normalization and baseline correction.**

**(a)** Raw spectrum $y$ vs fitted $w \cdot u - b$. **(b)** Trajectory of $u$ over iterations (blue → initial; red → final). **(c)** Convergence (log scale): residual $\|w \cdot u - b\|_2$ and total objective $\Phi(u, b)$. **(d)** Final $u$ normalized by its area (mean value to 1).

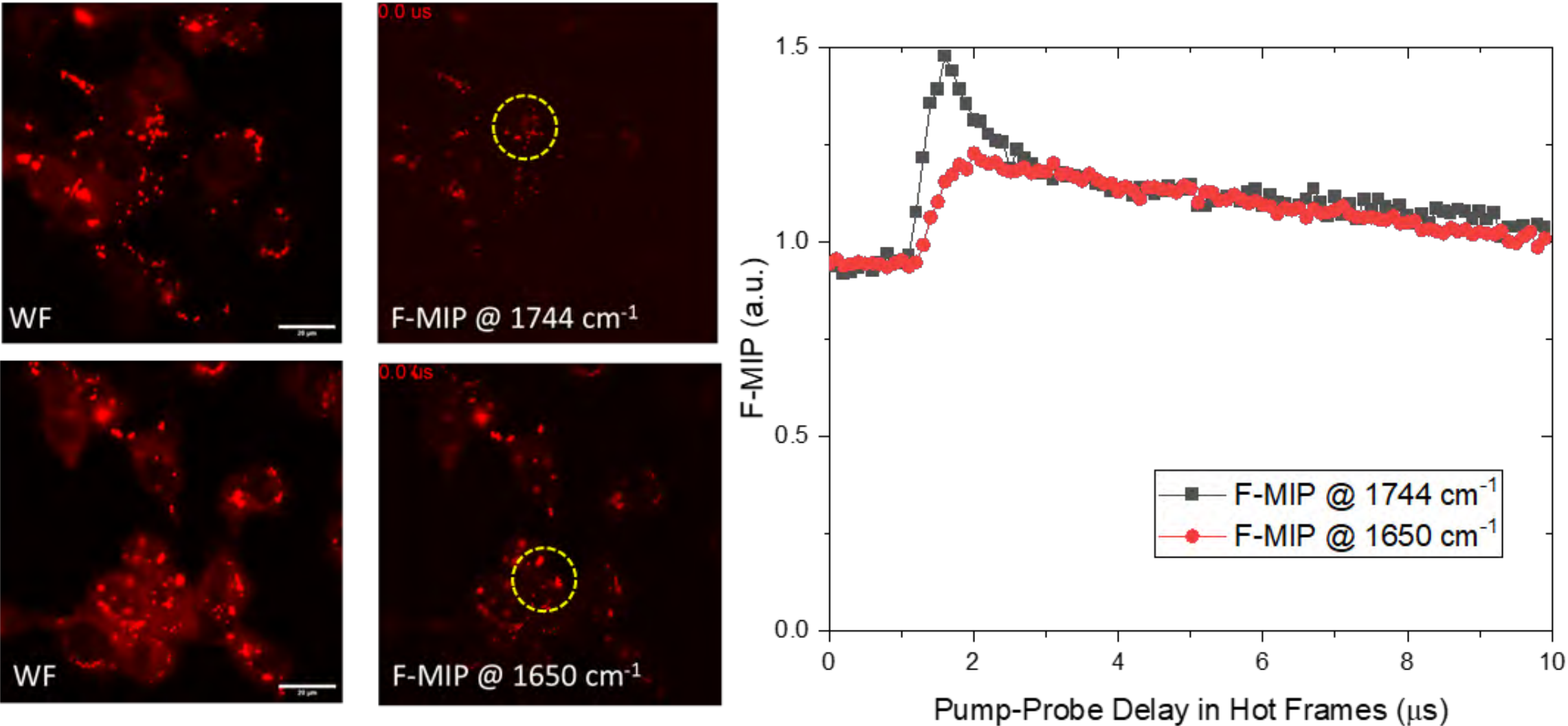

**Supplementary Figure 3 | Distinguishing transient thermal dynamics at C=O and water absorption bands.**

Wide-field (WF) and F-MIP fluorescence images of Hela cells with Lipi-red acquired at the lipid C=O stretch (1744 $cm^{-1}$, top row) and water absorption (1650 $cm^{-1}$, bottom row). Scale bars: 10 μm. The time-resolved F-MIP signal curves (right) show distinct thermal decay dynamics: the 1744 $cm^{-1}$ signal displays a sharper and faster rise, indicating localized lipid heating, whereas the 1650 $cm^{-1}$ signal reflects a broader, slower decay characteristic of water. Both curves exhibit a high baseline, attributed to environmental temperature difference due to the aqueous absorption.

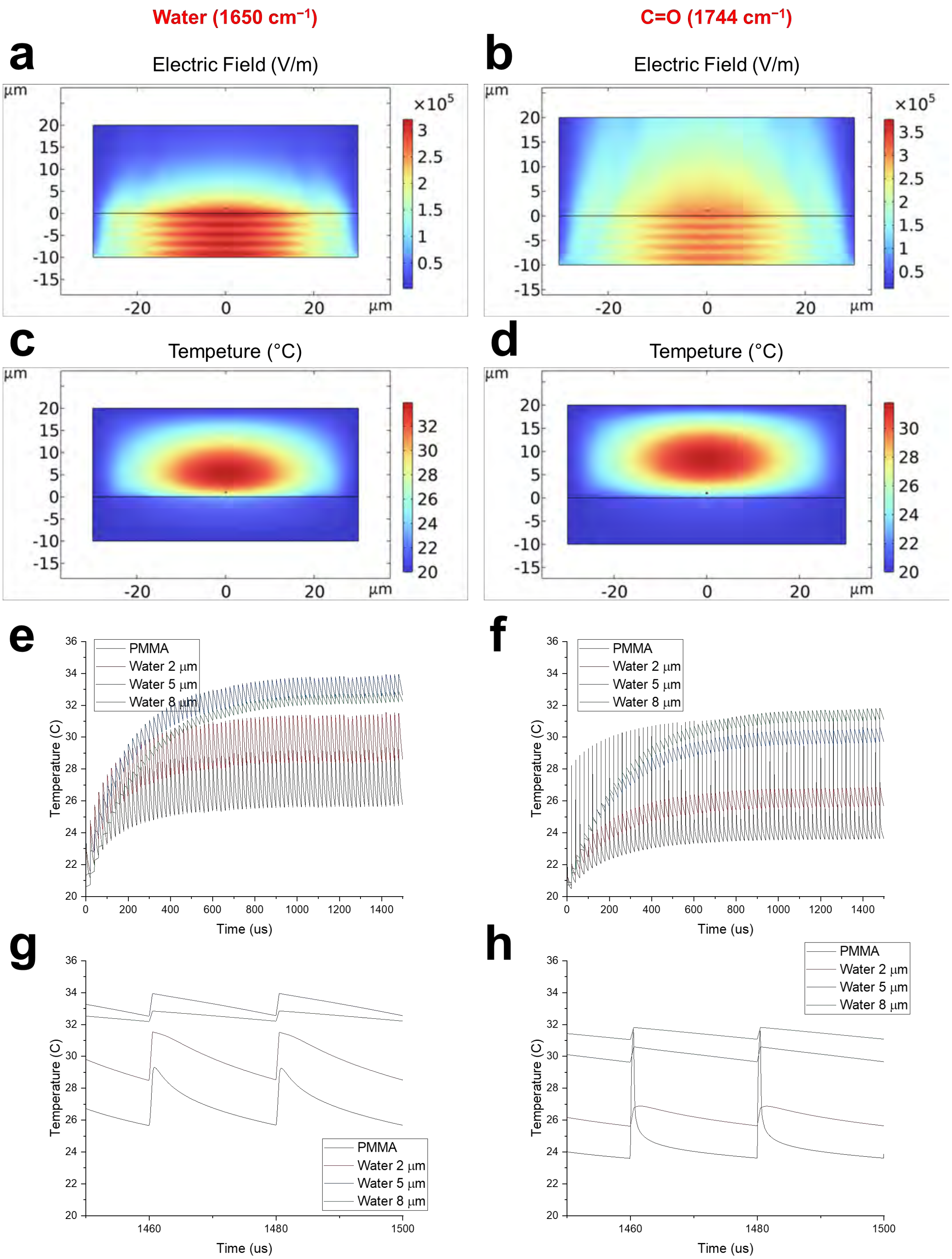

**Supplementary Figure 4 | COMSOL simulation of spatial thermal dynamics.**

COMSOL-simulated temperature distributions and time courses around a 100-nm PMMA bead in water following pulsed mid-IR absorption at two representative wavenumbers: 1650 $cm^{-1}$ (water absorption; left) and 1736 $cm^{-1}$ (PMMA C=O stretch; right). (a,b) Cross-sections of the transmitted mid-IR electric field. (c,d) Cross-sectional temperature maps at the end of the heating pulse, illustrating heat deposition and subsequent diffusion in the bead and surrounding medium. (e,f) Temperature dynamics over 0–1.5 ms measured at the bead center and in water at depths of 2, 5, and 8 μm from the water-$CaF_2$ interface. (g,h) Zoomed-in temperature dynamics over the final two pulse cycles (1.45–1.5 ms).

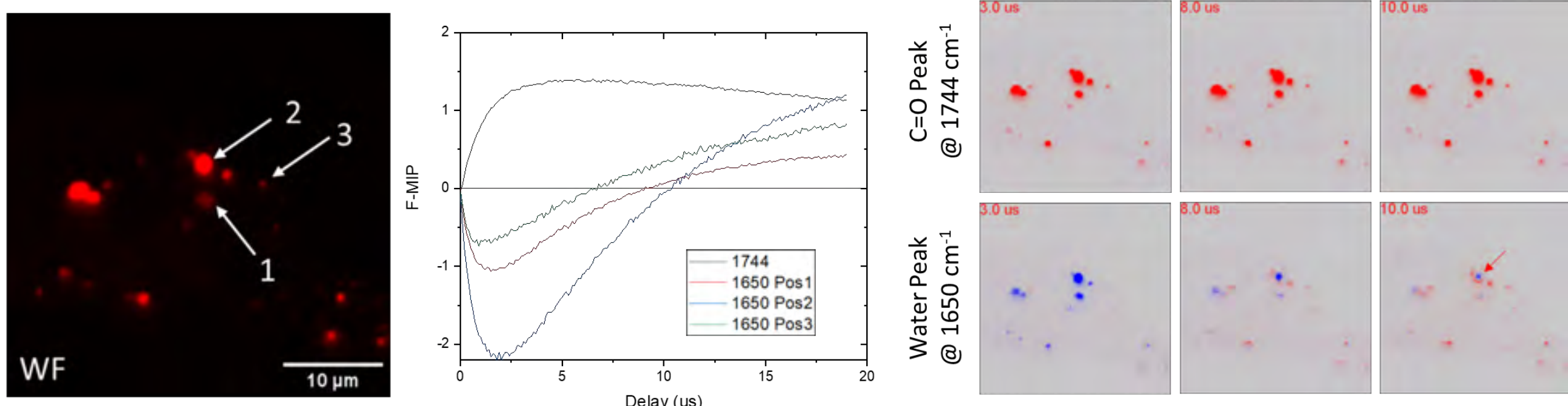

**Supplementary Figure 5 | Shape- and z-position–dependent transient thermal dynamics of lipid droplets.**

Left: Wide-field (WF) fluorescence image of intracellular lipid droplets with three representative droplets (Pos1–3) labeled. Middle: Time-resolved F-MIP traces acquired at 1650 $cm^{-1}$ for droplets at different positions (Pos1–3), compared to the reference signal at 1744 $cm^{-1}$. The decay curves show significant variation in temporal profiles, indicating depth-dependent thermal behavior. Right: Time-lapse differential images at 1744 $cm^{-1}$ (top row) and 1650 $cm^{-1}$ (bottom row), captured at 3.0, 8.0, and 10.0 µs delay. Notably, signals at 1650 $cm^{-1}$ exhibit polarity reversal and delayed rise consistent with shallower z-positions, confirming spatial dependence of photothermal signal dynamics. The lipid droplet indicated by the red arrow exhibits a clear outward-to-inward thermal propagation at 1650 $cm^{-1}$, consistent with heat diffusion through surrounding water.

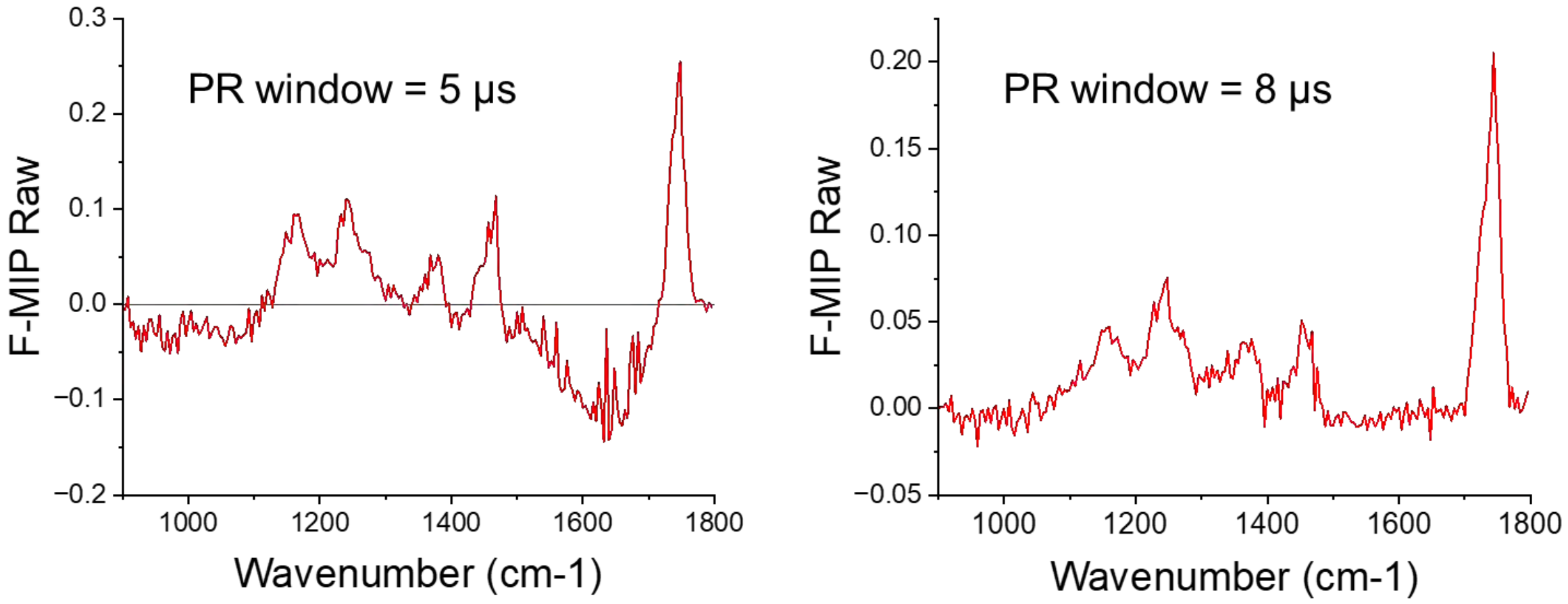

**Supplementary Figure 6 | Impact of photothermal relaxation window on F-MIP spectral contrast.**

F-MIP spectra of lipid droplets acquired using photothermal relaxation window of 5 μs (left) and 8 μs (right). The characteristic lipid vibrational features remain consistent across different settings, indicating that lipid-derived signals are stable over time. In contrast, the residual water absorption peak around 1650 $cm^{-1}$ shows significant variation in amplitude and shape depending on the window setting, reflecting the slow residual thermal decay of water. These results highlight that PR gating can selectively suppress background contributions without distorting the intrinsic spectral profile of the lipid signal.

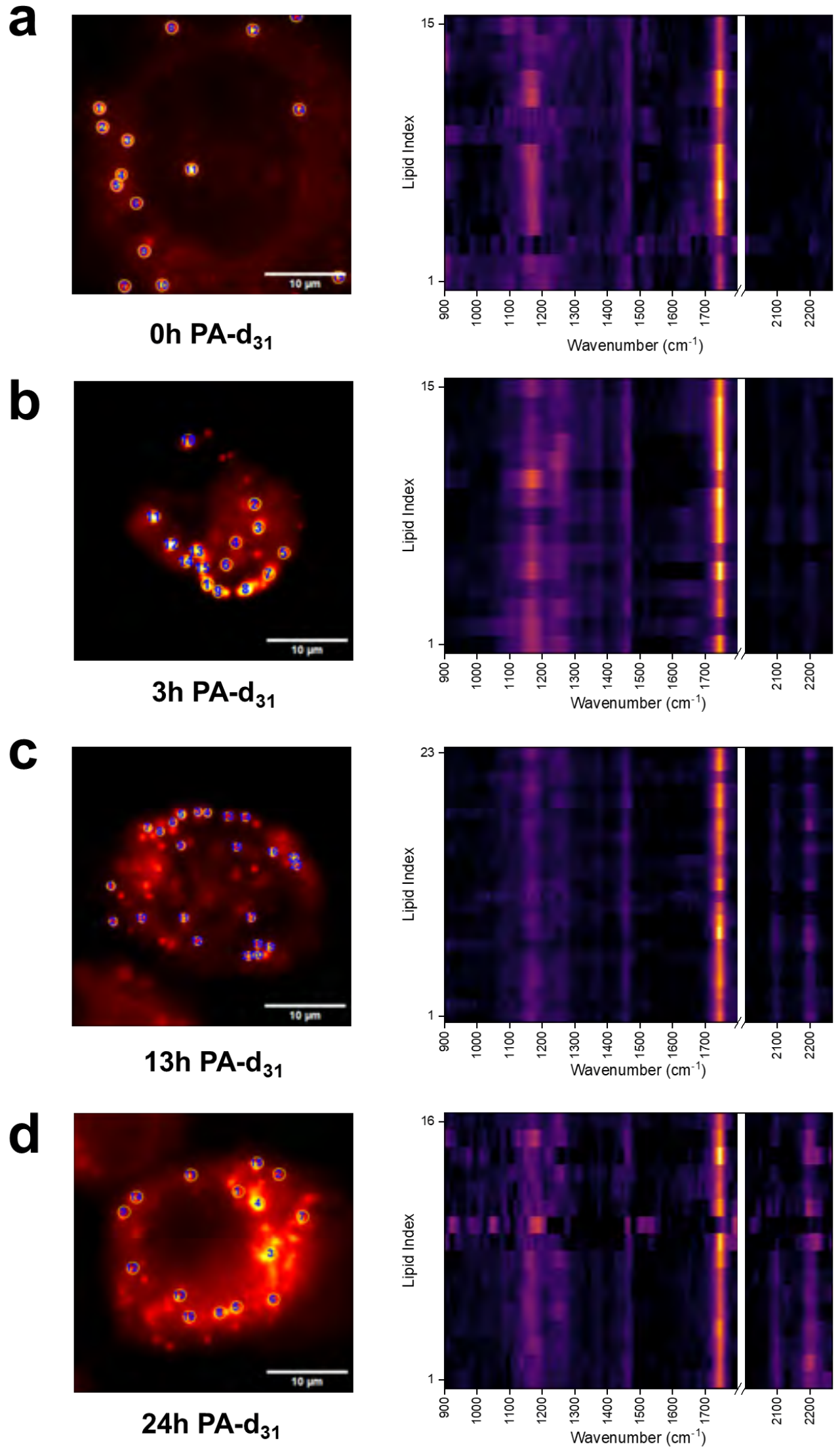

**Supplementary Figure 7 | Chem-SIM spectra of individual lipid droplets in OVCAR5 cells treated with fully-deuterated palmitic acid-$d_{31}$ (PA-$d_{31}$) for different time periods.**

Representative cells imaged after PA-d31 incubation duration of 0(a), 3(b), 13(c), and 24(d) hours. Left, Chem-SIM lipid-droplet maps with the analyzed droplets marked. Right, corresponding hyperspectral Chem-SIM signatures displayed as heat maps (each row represents one lipid droplet; color encodes the spectral intensity), spanning both the mid-IR fingerprint region and the C–D “silent window”. The ester C=O band (~1744 $cm^{-1}$) is observed across droplets, whereas C–D stretching features (2000–2300 $cm^{-1}$) emerge and strengthen with longer incubation, indicating progressive incorporation of PA-$d_{31}$ into neutral lipids. Scale bars, 10 µm.

Palmitic acid
(PA)

Palmitic acid-$d_{31}$
($PA-d_{31}$)

DAG + FA-CoA $\xrightarrow{\text{DGAT}}$ TAG

**Supplementary Figure 8 | Chemical structures of palmitic acid, palmitic acid-$d_{31}$, and cellular synthesis of TAG from DAG.**

Structures of palmitic acid (PA) and perdeuterated palmitic acid (PA-$d_{31}$), and a schematic of triacylglycerol (TAG) formation from diacylglycerol (DAG) via acylation with PA or PA-$d_{31}$ in cells, yielding deuterium-labeled TAG stored in lipid droplets. Deuterium atoms are indicated in blue.

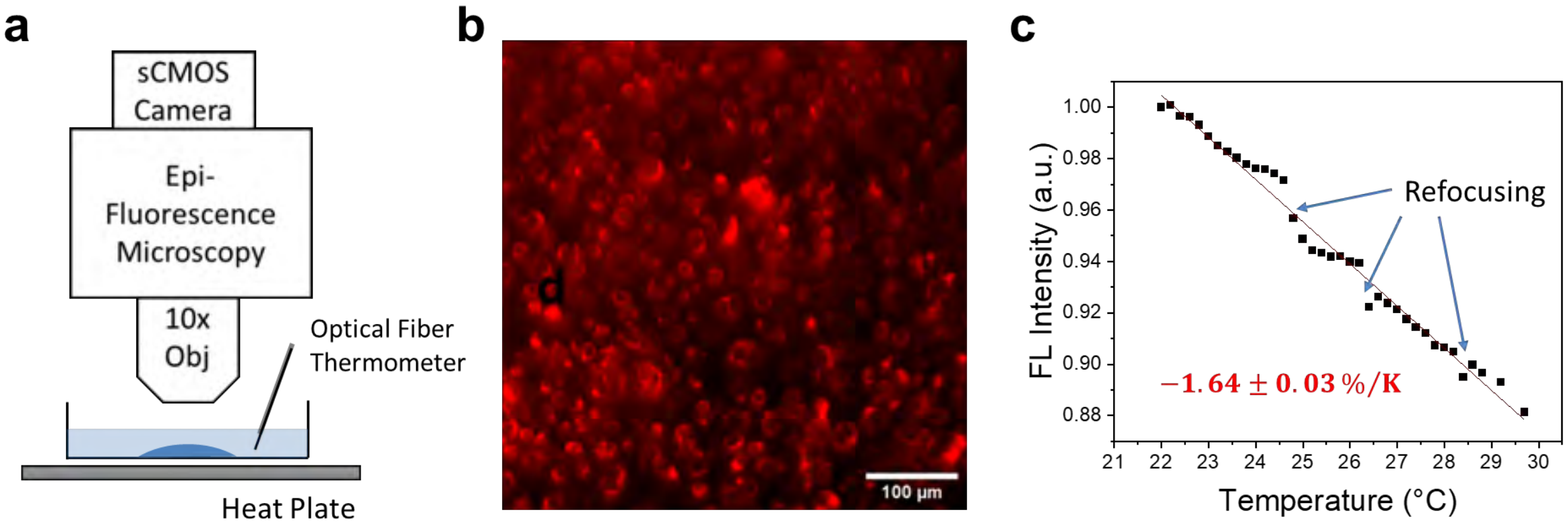

**Supplementary Figure 9 | Calibration of Lipi-Red fluorescence thermometry.**

(a) Schematic of the epifluorescence setup used to simultaneously record Lipi-Red fluorescence from fixed OVCAR5 cells and the culture-medium temperature with a fiber thermometer on a temperature-controlled stage. (b) Representative Lipi-Red fluorescence image; the total fluorescence intensity within the field of view is integrated for calibration. (c) Normalized Lipi-Red fluorescence intensity as a function of temperature, together with a linear fit, yielding a thermal sensitivity of 1.64±0.03 % $K^{-1}$.

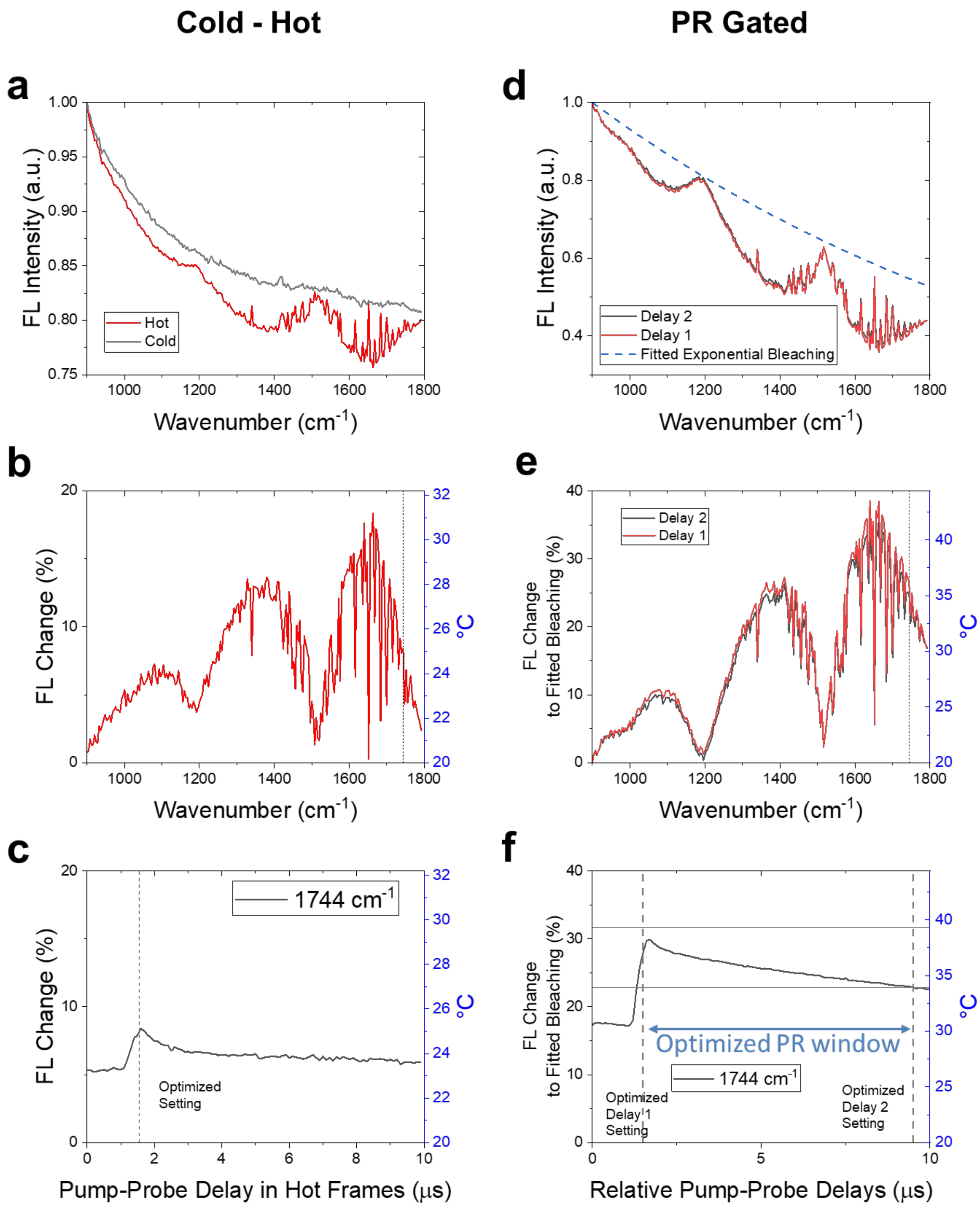

**Supplementary Figure 10 | Fluorescence thermometry of cell-culture medium under Chem-SIM mid-IR heating.**

(a) Representative fluorescence traces in Cold–Hot mode showing gradual photobleaching and an additional IR-induced drop in the Hot frames. (b) Cycle-averaged temperature of lipid droplets retrieved from the Cold–Hot modulation depth as a function of IR wavenumber; the temperature closely follows the on-sample IR power spectrum during the hyperspectral scan. (c) Transient temperature dynamics at the lipid C=O resonance in Cold–Hot mode, revealing distinct heating and cooling kinetics of the lipid droplets. (d) Representative fluorescence trace in photothermal relaxation (PR) mode, where delay-1 and delay-2 sample transient temperature rises superimposed on an exponential bleaching background (blue dashed fit). (e) Absolute temperatures extracted at the two PR delays versus wavenumber, showing nearly doubled temperature modulation compared with Cold–Hot mode. (f) Transient PR gated thermometry at C=O resonance yielding a peak temperature of 38.4 °C and a cycle-averaged temperature of 33.9 °C on lipid droplets (horizontal lines).

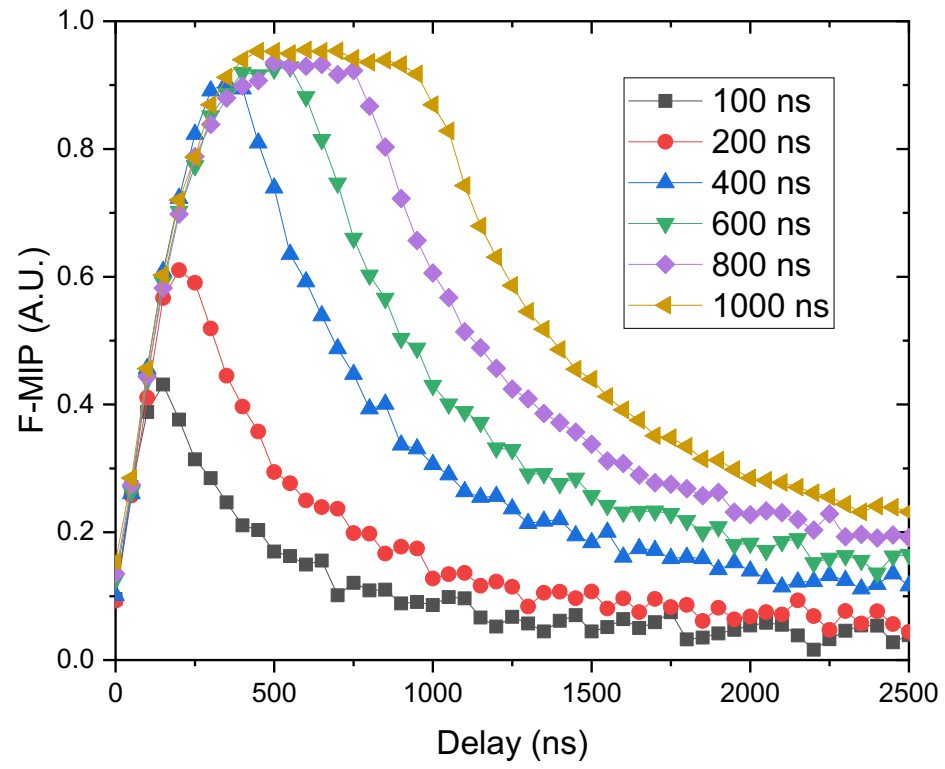

**Supplementary Figure 11 | F-MIP signal dynamics at various pump pulse widths.**

Measured with dried R6G labelled *S. aureus* sample at 1650 $cm^{-1}$.